\documentclass[reprint,superscriptaddress,a4paper,aps,showpacs,longbibliography]{revtex4-1}
\usepackage{graphicx}
\graphicspath{{./},{/home/user/fig/publi/muon/}}
\usepackage{amsfonts}
\usepackage{amsbsy}
\DeclareMathAlphabet\mathbfcal{OMS}{cmsy}{b}{n}
\usepackage{multirow}

\begin{document}

\title{Long-range dynamical magnetic order and spin tunneling in the cooperative paramagnetic states of the pyrochlore analogous spinel antiferromagnets CdYb$_2X_4$ ($X$~=~S, Se)}

\author{P. Dalmas de R\'eotier}
\author{C. Marin}
\author{A. Yaouanc}
\affiliation{Univ.\ Grenoble Alpes, CEA, INAC/PHELIQS, F-38000 Grenoble, France}
\author{C. Ritter}
\affiliation{Institut Laue-Langevin, Boite Postale 156X, F-38042 Grenoble Cedex 9, France}
\author{A.~Maisuradze}
\affiliation{Department of Physics, Tbilisi State University, Chavchavadze 3, 
GE-0128 Tbilisi, Georgia}
\author{B. Roessli}
\affiliation{Laboratory for Neutron Scattering and Imaging, Paul Scherrer Institute, 5232 Villigen-PSI, Switzerland}
\author{A. Bertin}
\affiliation{Univ.\ Grenoble Alpes, CEA, INAC/PHELIQS, F-38000 Grenoble, France}
\author{P.J. Baker}
\affiliation{ISIS Facility, STFC Rutherford Appleton Laboratory, Chilton, Didcot, OX11 0QX, UK}
\author{A. Amato}
\affiliation{Laboratory for Muon-Spin Spectroscopy, 
Paul Scherrer Institute, CH-5232 Villigen-PSI, Switzerland}

\date{\today}

\begin{abstract}

Magnetic systems with spins sitting on a lattice of corner sharing regular tetrahedra have been particularly prolific for the discovery of new magnetic states for the last two decades. The pyrochlore compounds have offered the playground for these studies, while little attention has been comparatively devoted to other compounds where the rare earth $R$ occupies the same sub-lattice, e.g.\ the spinel chalcogenides Cd$R_2X_4$ ($X = {\rm S}$, ${\rm Se}$).
Here we report measurements performed on powder samples of this series with $R$ = Yb using specific heat, magnetic susceptibility, neutron diffraction and muon-spin-relaxation measurements. The two compounds are found to be magnetically similar. They long-range order into structures described by the $\Gamma_5$ irreducible representation. The magnitude of the magnetic moment at low temperature is 0.77\,(1) and 0.62\,(1)\,$\mu_{\rm B}$ for $X = {\rm S}$ and ${\rm Se}$, respectively. Persistent spin dynamics is present in the ordered states. The spontaneous field at the muon site is anomalously small, suggesting magnetic moment fragmentation. A double spin-flip tunneling relaxation mechanism is suggested in the cooperative paramagnetic state up to 10~K. The magnetic space groups into which magnetic moments of systems of corner-sharing regular tetrahedra order are provided for a number of insulating compounds characterized by null propagation wavevectors. 

\end{abstract}

\maketitle

\section{Introduction}
\label{Intro} 

The study of geometrical frustration, where the crystal geometry prevents individual interactions from being satisfied, is one of the central themes in condensed matter research. Following the rich phase diagrams discovered for the insulating pyrochlore compounds  $R_2M_2$O$_7$, where $R$ is a trivalent  rare-earth ion and $M = {\rm Ti}, {\rm Sn}$ \cite{Gardner10,Gingras14}, there is a strong momentum to look at other compounds crystallizing within the same cubic crystal structure, but with different $M$ tetravalent elements. For example,  reports have been presented for $M = {\rm Zr}$ \cite{Lhotel15,Xu15,Xu16,Bonville16,Petit16,Wen17}, $M = {\rm Hf}$ \cite{Sibille16,Anand15,Anand16}, $M = {\rm Pb}$ \cite{Wiebe15,Hallas15}, $M = {\rm Pt}$ \cite{Cai16,Hallas16b,Li16}, and $M = {\rm Ge}$ \cite{Dun14,Dun15,Wiebe15,Li16,Hallas16a,Hallas16}.

\begin{figure}
\centering
\includegraphics[height=0.35\linewidth]{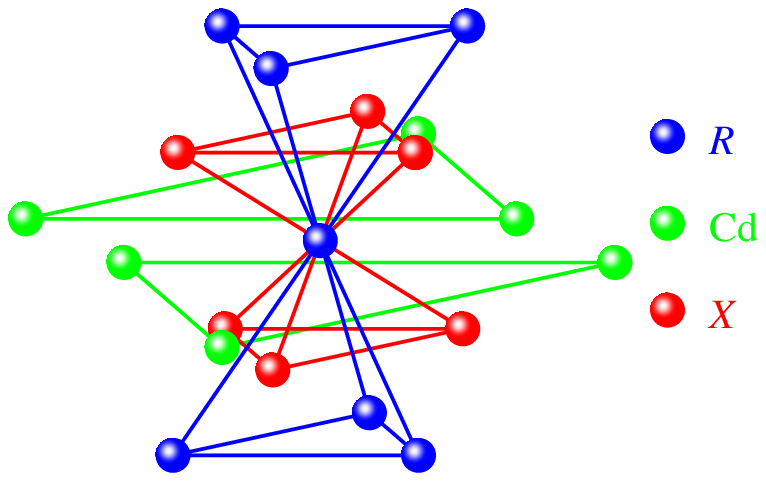}
\includegraphics[height=0.35\linewidth]{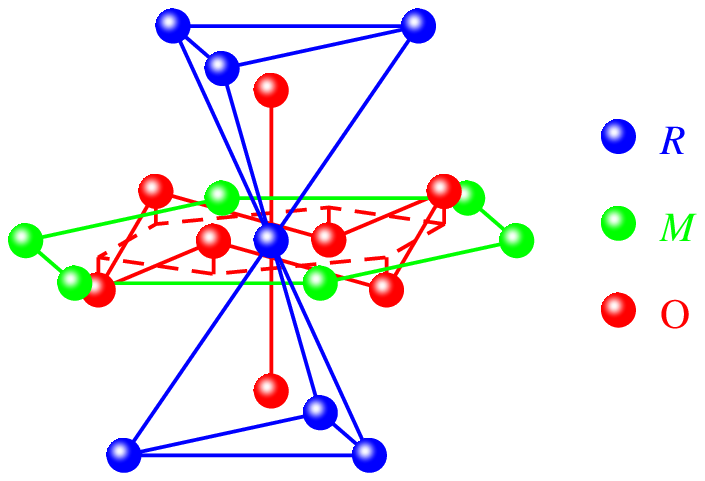}
\caption{(color online). 
Comparing the local environments at a rare-earth position for the spinel and pyrochlore compounds, on the left and right panels, respectively. Right-hand-side part of the figure reprinted with permission from Ref.~\onlinecite{Yaouanc11c}. Copyright 2011 by the American Physical Society. }
\label{structure}
\end{figure}

The crystal electric fields acting on the $R$ ions for these compounds \cite{Bertin12} are expected to be rather similar. This is understandable since the ion geometry arrangement around them is the same. Interestingly, there is another family of compounds for which the $R$ ions also sit on a lattice of corner-sharing regular tetrahedra: the cadmium chalcogenide spinels \cite{Lau05}. They are of particular interest because the point symmetry ($\overline{3}m$) at the $R$ site is the same as in the pyrochlores but the coordination and bonding around $R$ with the O$^{2-}$ neighboring ions forming a nearly regular octahedron is very different; see Fig.~\ref{structure}.
Therefore their crystal-electric fields (CEFs) should be substantially different. This is confirmed by the appearance of a spin-ice state in CdEr$_2$Se$_4$ \cite{Lago10} and  possibly in CdEr$_2$S$_4$ \cite{Legros15}, and its absence in CdHo$_2$S$_4$ \cite{Yaouanc15}. This is in contrast to the pyrochlore titanates or stannates for which the spin ice state is observed when $R$ = Ho, whereas the $R$ = Er systems have planar local anisotropy. Therefore the spinels offer the possibility to extend the study of the lattice of corner sharing $R$ tetrahedra to unknown territory thanks to different CEFs.

Here, we report an investigation of the insulating spinel chalcogenides CdYb$_2X_4$ with $X = {\rm S}$ or ${\rm Se}$ performed on powder samples using specific heat, magnetic susceptibility, neutron diffraction and muon-spin-relaxation ($\mu$SR) measurements. Compared to the study performed by Higo and collaborators for the same compounds \cite{Higo17}, we determine their magnetic structure and present a characterization of their spin dynamics using $\mu$SR. In contrast to the pyrocholore ytterbium stannate and titanate which are splayed ferromagnets \cite{Yaouanc13,Yaouanc16}, the two ytterbium spinels are antiferromagnets with magnetic moments perpendicular to their local three-fold axis. However, the observed paramagnetic spin tunneling is rather similar to previous findings for Yb$_2$Ti$_2$O$_7$, Yb$_2$Sn$_2$O$_7$ and Nd$_2$Sn$_2$O$_7$ \cite{Dalmas16a, Dalmas17a}, pointing to the topology of the corner-sharing tetrahedra lattice as its origin.

\section{Experimental}
\label{Experimental}

The synthesis of CdYb$_2$S$_4$ and CdYb$_2$Se$_4$ was achieved in a two-step approach using sealed quartz ampoules. High purity starting materials (5N) were used: ytterbium metal, sulfur, and CdS, or selenium and CdSe. The first step was the preparation of Yb$_2$S$_3$ and Yb$_2$Se$_3$ by heat treatment up to $650^\circ$C. The second step consisted of grinding the mixture CdS and Yb$_2$S$_3$ (or CdSe and Yb$_2$Se$_3$), then pressing it into pellets. The final solid state reaction leading to CdYb$_2$S$_4$ and CdYb$_2$Se$_4$ was achieved by heating the pellets up to $900^\circ$C for two weeks. The single phase nature of the obtained compounds was checked by X-ray powder diffraction. A more detailed discussion of the crystalline purity of the two samples will be presented in Sec.~\ref{microscopic_neutron}.

The heat capacity and susceptibility measurements were performed with commercial instruments, namely the Physical Property Measurement System and the Magnetic Property Measurement System, both from Quantum Design, inc. The neutron diffraction measurements were conducted at the D20 high-intensity powder two-axis diffractometer of the Institut Laue Langevin, Grenoble, France. The muon spin rotation and relaxation measurements ($\mu$SR) were mostly performed at the MuSR spectrometer of the ISIS pulsed muon source, Rutherford Appleton Laboratory, Chilton, UK, and partly at the GPS spectrometer of the Swiss Muon Source, Paul Scherrer Institute, Villigen, Switzerland.

Owing to the strong neutron absorption cross-section of Cd, we used a hollow cylinder sample holder for the diffraction experiments: for each compound $\approx 8$~g of powder were inserted into the space available between the two coaxial cylinders of diameter 14 and 16~mm. 

\section{Results}
\label{Results}

\subsection{Macroscopic measurements}
\label{macroscopic}

\begin{figure}
{\includegraphics[width=0.9\linewidth ]{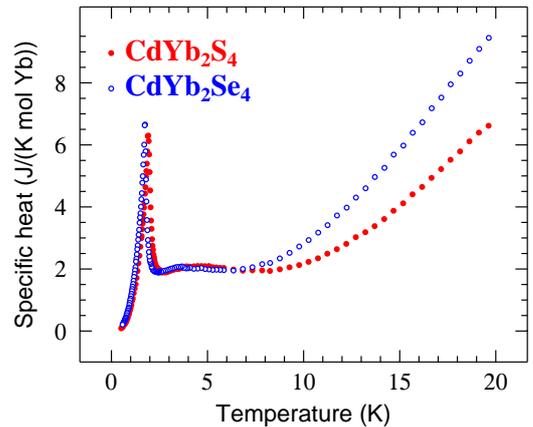}}
\caption{(color online). 
Heat capacity of CdYb$_2$S$_4$ and CdYb$_2$Se$_4$ versus temperature.}
\label{CdYb2X4_sh}
\end{figure}

In Fig.~\ref{CdYb2X4_sh} we display the specific heat $C_{\rm p}$ in the whole temperature range.
 The results for the two compounds are rather similar, except above $\approx 7$~K for which the thermal response is larger for CdYb$_2$Se$_4$. This is essentially explained by a sizably larger molar mass of this system which shifts down its phonon spectrum relative to the other. This will not concern us here. Similar peaks are observed just below 2~K pointing out thermodynamic phase transitions. The peak shape suggests these transitions to be second order. We shall determine with neutron and $\mu$SR that they are of magnetic origin. Only mild bumps in $C_{\rm p}$  are observed between 2 and 6~K. Hence, short-range magnetic correlations in the correlated states are relatively weak in these compounds \cite{Dalmas03}.

\begin{figure}
{\includegraphics[width=\linewidth ]{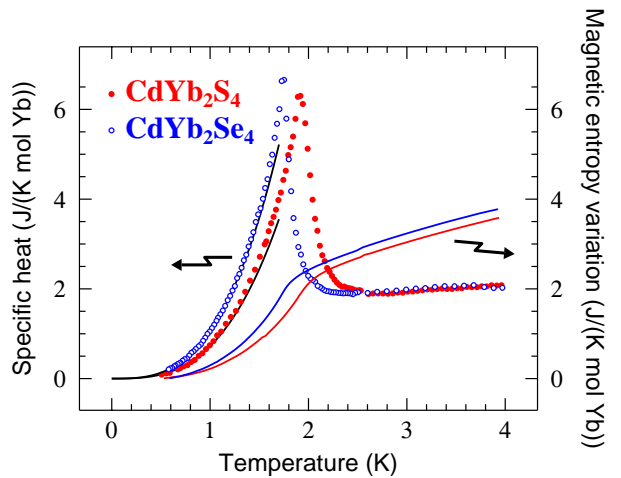}}
\caption{(color online). 
Low temperature part of Fig.~\ref{CdYb2X4_sh}. The black full lines result from a fit of the function $C_{\rm p} = \mathcal{B}T^3$ to the data below 1.2~K, with $\mathcal{B}$ = 0.73\,(3) and 1.06\,(5)~J\,K$^{-4}$\,mol$^{-1}$ for CdYb$_2$S$_4$ and CdYb$_2$Se$_4$ respectively. The color full lines display the variation of magnetic entropy deduced from $C_{\rm p}(T)$.}
\label{CdYb2X4_sh_zoom}
\end{figure}

In Fig.~\ref{CdYb2X4_sh_zoom} the low temperature parts of $C_{\rm p}$ are shown. 
The critical temperatures $T_{\rm c}$ extracted from $C_{\rm p}$ are listed in Table~\ref{material_parameters}. 
\begin{table}
\caption{Different physical parameters resulting from this study. The quantities are the spin wave velocity $v_{\rm sw}$, the lattice parameter $a$, the position parameter $x$ for the chalcogen element (determined at 10 and 2.7~K for the sulfide and selenide, respectively), the critical temperature $T_{\rm c}$ of the magnetic transition derived from the heat capacity data, and the spontaneous moments at 100~mK. The units are given in the second row. The uncertainties are only of statistical origin.
  }
\label{material_parameters}
\resizebox{\linewidth}{!}{%
\begin{tabular}{lccccc}
\hline\hline
             & $v_{\rm sw}$         & $a$ & $x$ & $T_{\rm c}$ & $m(T=100~{\rm mK})$\cr
             & m\,s$^{-1}$ & \AA & -   & K          & $\mu_{\rm B}$ \cr
\hline
CdYb$_2$S$_4$  & 141\,(2) & 11.003\,(2) & 0.2594\,(5) & 1.92 & 0.77\,(1) \cr
CdYb$_2$Se$_4$ & 130\,(2) & 11.455\,(2) & 0.2575\,(3) & 1.75 & 0.62\,(1) \cr
\hline\hline
\end{tabular}
}
\end{table}
Below $T_{\rm c}$ we find $C_{\rm p} \propto T^3$, a power law behavior expected for a conventional antiferromagnet. The spin wave velocities $v_{\rm sw}$, deduced from $C_{\rm p}$ following the method explained elsewhere \cite{Dalmas12a}, are given in Table~\ref{material_parameters}. Knowing $v_{\rm sw}$, the scale of the exchange integral  ${\mathcal I}$ can be inferred \cite{Dalmas12a}. We get ${\mathcal I}/k_{\rm B}$ = 0.40\,(1)\,K and 0.35\,(1)\,K for the sulfide and the selenide, respectively. Here, $k_{\rm B}$ is Boltzmann's constant.
In line with the similarity of $C_{\rm p}$, the variation of magnetic entropy (Fig.~\ref{CdYb2X4_sh_zoom}) follows the same trend for the two compounds, except for a temperature shift corresponding to the transition temperature difference. Between 0.5 and 4~K this variation approaches 4~J/(K mol Yb), i.e. $\approx 0.7\,R \ln 2$ where $R$ is the ideal gas constant.

\begin{figure}
\includegraphics[width=0.9\linewidth]{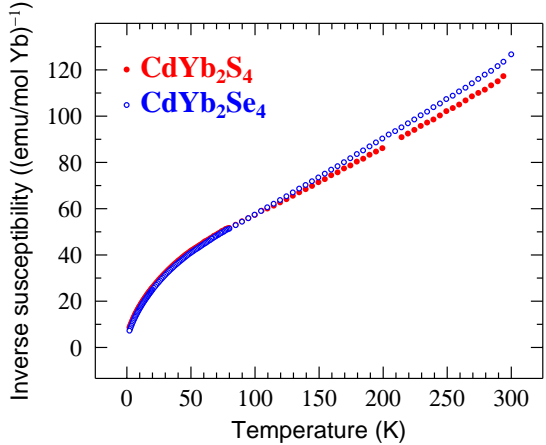}
\caption{(color online). 
  Inverse susceptibility versus temperature for CdYb$_2$S$_4$ and CdYb$_2$Se$_4$. The data have been recorded with an applied field of 0.9 and 1.5~mT, respectively. The fields are sufficiently low for the data to reflect the susceptibilities.}
\label{CdYb2X4_inv_chi}
\end{figure}

The inverse susceptibility data are collected in Fig.~\ref{CdYb2X4_inv_chi}.
The magnetic responses are rather similar, in particular at low temperatures. The expected straight line in a Curie-Weiss picture can be approximately found above $\approx 50$~K and at low temperature. However the Curie-Weiss temperature $\theta_{\rm CW}$ deduced from such analysis would depend on the temperature range at which the analysis is performed. In fact, the thermal behavior of the susceptibilities is strongly influenced by the crystal-fields acting at the rare-earth ions \cite{Higo17}. We will therefore use the mean-field approximation formula $\mathcal{I}$ = $3k_{\rm B}|\theta_{\rm CW}|/z_{\rm nn}J(J+1)$ (see, e.g.\ Ref.~\cite{Dalmas12a}) to deduce $|\theta_{\rm CW}| \approx 13$ and 11~K for the sulfide and selenide, respectively. Here $J$ = 7/2 is the spin of the Yb$^{3+}$ manifold and $z_{\rm nn}$ = 6 is the number of nearest neighbors to a magnetic ion. The $\theta_{\rm CW}$ values are consistent with those found by Higo {\em et al}. With a ratio $f = |\theta_{\rm CW}|/T_{\rm c} \gtrsim 6$ for the frustration index \cite{Ramirez01}, we conclude to the influence of frustration on the magnetic properties of these two spinels. 

Our bulk data clearly show that the two compounds have similar magnetic properties. Table~\ref{material_parameters} suggests the selenide to be slightly less magnetic than the sulfide.

We now turn to the microscopic probe measurements which reveal exotic magnetic properties.

\subsection{Microscopic probe measurements}
\label{microscopic}

\subsubsection{Neutron diffraction results}
\label{microscopic_neutron}

The crystallographic structure of the CdYb$_2$S$_4$ and CdYb$_2$Se$_4$ spinels is described according to the Fd$\bar{3}$m cubic space group \cite{Ben-Dor80,Suchow64}. With the adopted description where the origin of the cubic unit cell is at a point of symmetry $\bar{3}$m, the Cd and Yb ions respectively occupy 8a and 16d Wyckoff positions of respective coordinates ($\frac{1}{8}$, $\frac{1}{8}$, $\frac{1}{8}$) and ($\frac{1}{2}$, $\frac{1}{2}$, $\frac{1}{2}$). The S or Se ions are located at a 32e position ($x$, $x$, $x$). From our Rietveld neutron refinements using Fullprof \cite{Rodriguez93}, we obtain the lattice parameters $a$ and the chalcogen position parameters $x$ as given in Table~\ref{material_parameters}.

\begin{figure}
{\includegraphics[width=0.9\linewidth ]{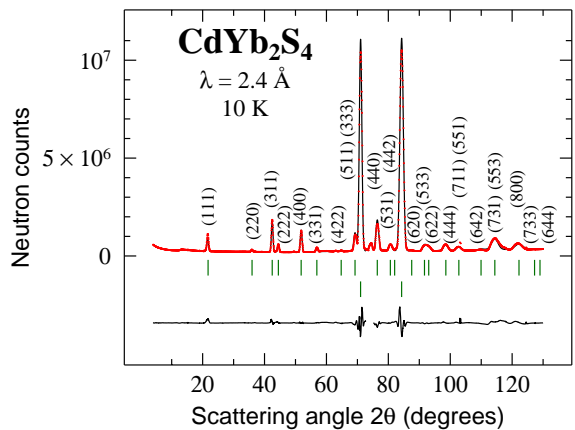}}
\caption{(color online). 
Neutron diffraction pattern for a powder of CdYb$_2$S$_4$ at 10~K versus the scattering angle $2\theta$. Neutrons of wavelength 2.4~\AA\ were used. The experimental data are drawn as red dots. The solid line shows the result of a Rietveld refinement. The bottom line displays the difference between the experimental data and the refinement. The observed reflections are labeled with Miller indices. The vertical markers indicate the position of the expected Bragg peaks. The second row of markers correspond to Bragg peaks from the Cu sample container.}
\label{CdYb2S4_nuclear}
\end{figure}

\begin{figure}
{\includegraphics[width=0.9\linewidth ]{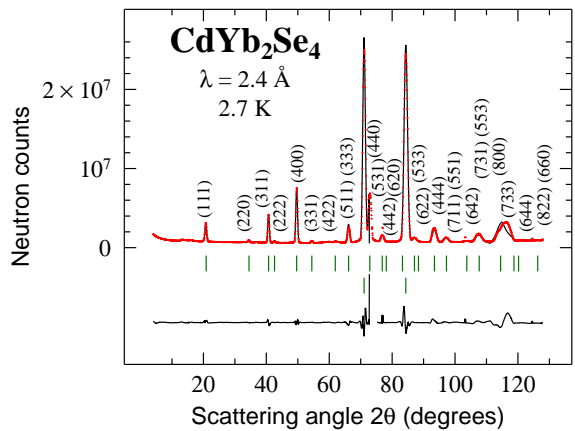}}
\caption{(color online). 
Neutron diffraction pattern for a powder of CdYb$_2$Se$_4$ at 2.7~K versus the scattering angle $2\theta$. Neutrons of wavelength 2.4~\AA\ were used. The experimental data are drawn as red dots. The solid line shows the result of a Rietveld refinement. The bottom line displays the difference between the experimental data and the refinement. The observed reflections are labeled with Miller indices. The vertical markers indicate the position of the expected Bragg peaks. The second row of markers correspond to Bragg peaks from the Cu sample container.}
\label{CdYb2Se4_nuclear}
\end{figure}

Figures~\ref{CdYb2S4_nuclear} and \ref{CdYb2Se4_nuclear} display the diffraction diagrams recorded at 10~K for CdYb$_2$S$_4$ and 2.7~K for CdYb$_2$Se$_4$, respectively. For both compounds the width of the Bragg peaks corresponds to the instrument resolution. The structure refinements are very good. Only a peak at $\approx 74.5^\circ$ and perhaps two very tiny contributions at about 63$^\circ$ and 97$^\circ$ in the CdYb$_2$S$_4$ data are not present in the model, suggesting the presence of a minority parasitic phase in this sample.

\begin{figure}
{\includegraphics[width=0.9\linewidth ]{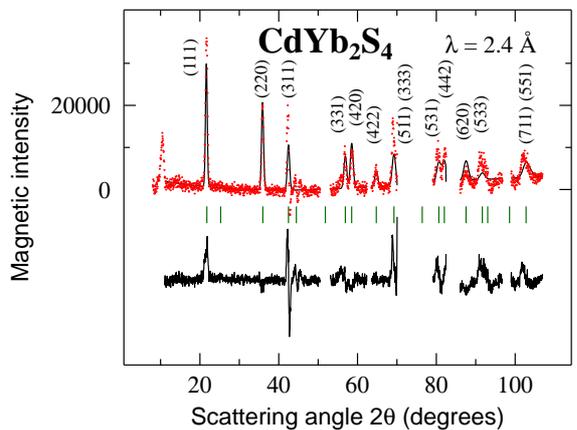}}
\caption{(color online). 
Magnetic neutron scattering pattern of CdYb$_2$S$_4$, i.e.\ difference between the diagrams recorded at 100~mK and 10~K. The fit according to the magnetic structure described in the main text is shown as a solid line. The bottom line displays the difference between the experimental data and the refinement. The vertical markers indicate the positions of the possible magnetic Bragg peaks while the observed reflections are labeled with Miller indices.
}
\label{CdYb2S4_magnetic}
\end{figure}

\begin{figure}
{\includegraphics[width=0.9\linewidth ]{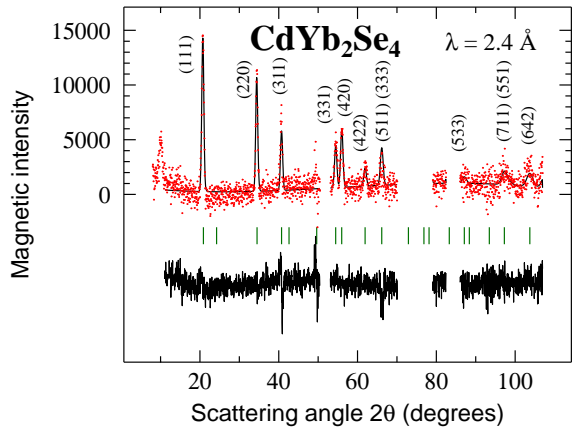}}
\caption{(color online). 
Magnetic neutron scattering pattern of CdYb$_2$Se$_4$, i.e.\ difference between the diagrams recorded at 100~mK and 2.7~K. The fit according to the magnetic structure described in the main text is shown as a solid line. The bottom line displays the difference between the experimental data and the refinement. The vertical markers indicate the positions of the possible magnetic Bragg peaks while the observed reflections are labeled with Miller indices.
}
\label{CdYb2Se4_magnetic}
\end{figure}

The magnetic scattering data deduced from the difference of the diffraction diagrams recorded at 100~mK and 10~K for CdYb$_2$S$_4$ and at 100~mK and 2.7~K for CdYb$_2$Se$_4$ exhibit neutron intensity at the positions of the nuclear Bragg peaks. An additional narrow peak is observed at scattering angle $2\theta \approx$ 10.3$^\circ$ and 10.1$^\circ$ for the sulfide and selenide, respectively (Figs.~\ref{CdYb2S4_magnetic} and \ref{CdYb2Se4_magnetic}). 
Relative to the other magnetic peaks, its intensity is somewhat stronger for the former compound. Remarkably, the peak observed in CdYb$_2$Se$_4$ could be indexed as ($\frac{1}{2}$, $\frac{1}{2}$, $\frac{1}{2}$). This is not the case for the sulfide peak for which the ($\frac{1}{2}$, $\frac{1}{2}$, $\frac{1}{2}$) peak would be shifted by 0.3-0.4$^\circ$ compared to the observed peak. It would be tempting to assign these peaks at low angles to a structure with a magnetic propagation vector  equal or close to ${\bf k}$ = $(\frac{1}{2}$, $\frac{1}{2}$, $\frac{1}{2})$. This idea suffers, however, from two strong reservations: (i) there are no further magnetic peaks which could be linked to this hypothetical propagation vector and (ii) some scattering intensity is still observed around 10$^\circ$ in CdYb$_2$S$_4$ at 2.5~K, i.e.\ above the $T_{\rm c}$ value deduced from the heat capacity measurements and at a temperature where all other magnetic peaks have vanished. These facts strongly suggest that the intensity observed around 10$^\circ$ in both compounds arises from a long-range ordered magnetic parasitic phase present in the two compounds, a nuclear reflection of which is observed at 10~K and 74.5$^\circ$ in the sulfide. For the analysis of the magnetic scattering data, only the intensity recorded above 11$^\circ$ for the two systems will be considered.

\begin{figure}
  {\includegraphics[width=\linewidth ]{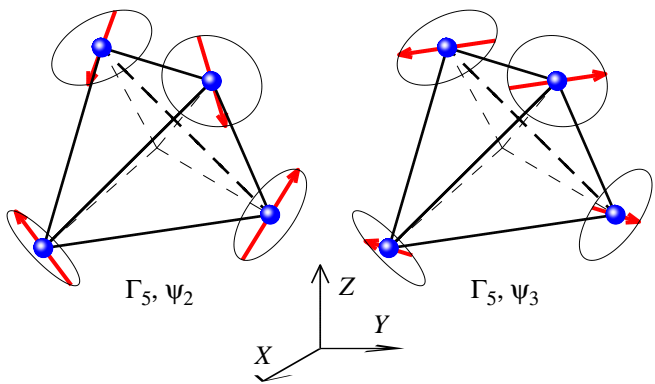}}
\caption{(color online). Magnetic structures corresponding to the $\psi_2$ (left) and $\psi_3$ (right) modes of the $\Gamma_5$ irrep. The thin dashed lines represent the three-fold local symmetry axes. For both structures, the magnetic moments are confined to planes perpendicular to these axes. The cubic axes are denoted as $(X,Y,Z)$.
}
\label{structure_psi2_psi3}
\end{figure}

The magnetic diffraction patterns are displayed in Figs.~\ref{CdYb2S4_magnetic} and \ref{CdYb2Se4_magnetic}%
for CdYb$_2$S$_4$ and CdYb$_2$Se$_4$ respectively. Following the previous discussion, since the magnetic intensity of the CdYb$_2$S$_4$ and CdYb$_2$Se$_4$ phases is only observed at the nuclear Bragg peak positions, we conclude that the magnetic structure propagation wavevector is ${\bf k}$ = (0, 0, 0). The magnetic order is long range since the width of the magnetic Bragg peaks corresponds to the diffractometer resolution. Following a representation analysis study for the symmetry at the rare earth site, the possible magnetic structures are obtained from linear combinations of the basis vectors belonging to one of the four irreducible representations (irreps) labeled $\Gamma_{3,5,7,9}$ (see, e.g.\ Ref.~\onlinecite{Wills06}). From the presence of magnetic scattering for both systems at the (111) and (220) Bragg positions and its absence at (200), the $\Gamma_{3,7,9}$ irreps are excluded. Satisfactory Rietveld refinements of the experimental data (Figs.~\ref{CdYb2S4_magnetic} and \ref{CdYb2Se4_magnetic}) are obtained with a magnetic structure belonging to the $\Gamma_5$ representation. This representation is two-dimensional with basis vectors denoted as $\psi_2$ and $\psi_3$. The magnetic structures arising from $\psi_2$, $\psi_3$, or any linear combination of them yield the same neutron diffraction pattern for a powder. For a reliable selection, measurements with polarized neutrons on crystals would be needed \cite{Poole07}. Powder measurements under an applied magnetic field are an alternative \cite{Champion03}. The common feature of $\psi_2$ and $\psi_3$ is that the magnetic moment at each rare earth is perpendicular to the local trigonal axis (Fig.~\ref{structure_psi2_psi3}).

\begin{figure}
{\includegraphics[width=0.9\linewidth ]{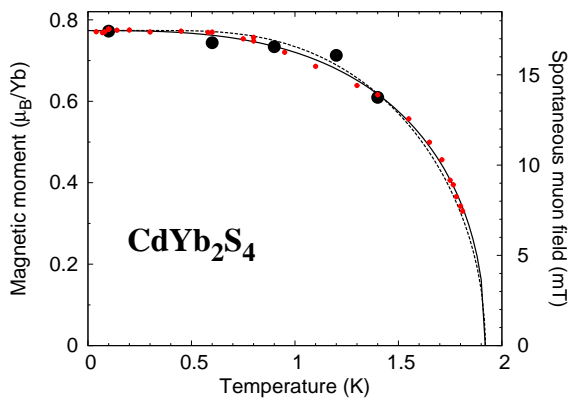}}
\caption{(color online). Temperature dependence of the Yb$^{3+}$ magnetic moment determined from neutron scattering (large black bullets) and the spontaneous field at the muon site (small red bullets) measured for CdYb$_2$S$_4$. The error bars on the experimental data are smaller or equal to the symbol size. The full line is a fit of the phenomenological law $B_0[1-(T/T_{\rm c})^\alpha]^\beta$ to the spontaneous field. The parameters are $B_0$ = 17.46\,(4)~mT, $\alpha$ = 2.93\,(13) and $\beta$ = 0.447\,(14). The dashed line shows the prediction of a mean-field model based on the $S =1/2$ Brillouin function.
}
\label{CdYb2S4_moment}
\end{figure}

\begin{figure}
{\includegraphics[width=0.9\linewidth ]{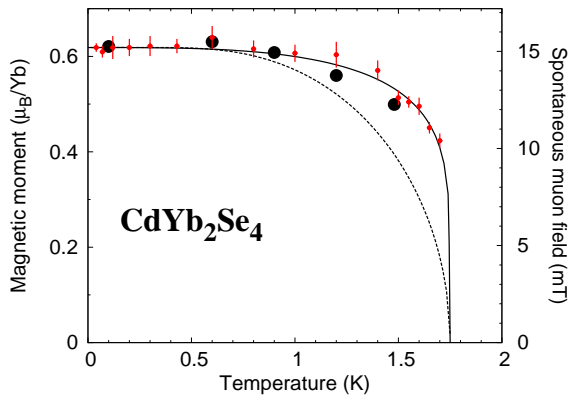}}
\caption{(color online). Temperature dependence of the Yb$^{3+}$ magnetic moment determined from neutron scattering (large black bullets) and the spontaneous field at the muon site (small red bullets) measured for CdYb$_2$Se$_4$. The error bars on the magnetic moment are smaller or equal to the symbol size. The full line is a fit of the same phenomenological law  as in Fig.~\ref{CdYb2S4_moment} to the spontaneous field with $B_0$ = 15.23\,(9)~mT, $\alpha$ = 3.2\,(6) and $\beta$ = 0.17\,(2). The dashed line shows the prediction of a mean-field model based on the $S =1/2$ Brillouin function.
}
\label{CdYb2Se4_moment}
\end{figure}

Figures~\ref{CdYb2S4_moment} and \ref{CdYb2Se4_moment} display the temperature dependences of the Yb$^{3+}$ magnetic moment as determined by neutron diffraction. 
The magnetic moments at 100~mK are listed in Table~\ref{material_parameters}. The magnetic moment moduli are somewhat reduced relative to $\approx 1\ \mu_{\rm B}$ in the Yb$_2$Ti$_2$O$_7$ and Yb$_2$Sn$_2$O$_7$ pyrochlore compounds, illustrating the difference between the pyrochlores and the spinels. An understanding of this difference requires to determine the crystal-field ground state and the strength of the molecular field at the rare-earth site.

\subsubsection{$\mu$SR results}
\label{microscopic_muon}

\begin{figure}
\includegraphics[width=0.9\linewidth ]{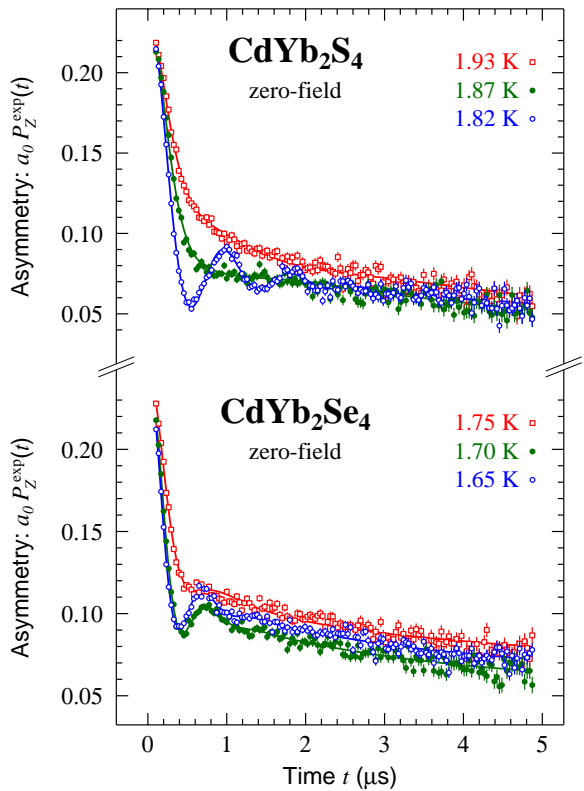}
\caption{(color online). 
Zero-field $\mu$SR spectra recorded close to the magnetic transition in CdYb$_2$S$_4$ (top) and CdYb$_2$Se$_4$ (bottom). The oscillations appear in a very narrow temperature range and are not overdamped. This is an indication for the good quality of the samples. It also suggests that critical dynamics does not play an important role.}
\label{CdYb2S4_muSR}
\end{figure}

In Fig.~\ref{CdYb2S4_muSR} we display asymmetry spectra recorded close to the magnetic transition for the two compounds.
The detection of oscillations characteristic of spontaneous muon precession means that a compound is magnetically ordered at the temperature of observation \cite{Yaouanc11}. It is clearly the case at 1.70 and 1.65~K for the selenide and 1.82~K for the sulfide. The $T_{\rm c}$ values from specific heat and the measured $\mu$SR spectra are consistent. The spontaneous field at the muon site is plotted versus temperature in Figs.~\ref{CdYb2S4_moment} and \ref{CdYb2Se4_moment}, respectively for the two compounds. The  thermal variations deduced from neutron diffraction and $\mu$SR are consistent. 

To conclude our discussion of the spectra below $T_{\rm c}$, we note the finite slopes at long time. They are in fact temperature independent down to far below 0.1~K. The associated spin-lattice relaxation rate is $\lambda_Z \approx 0.4 \, \mu {\rm s}^{-1}$ for the two compounds, a signature of the so-called persistent spin dynamics \cite{Dalmas16a}.

\begin{figure}
\begin{picture}(255,434)
\put(0,270)
{\includegraphics[width=0.9\linewidth ]{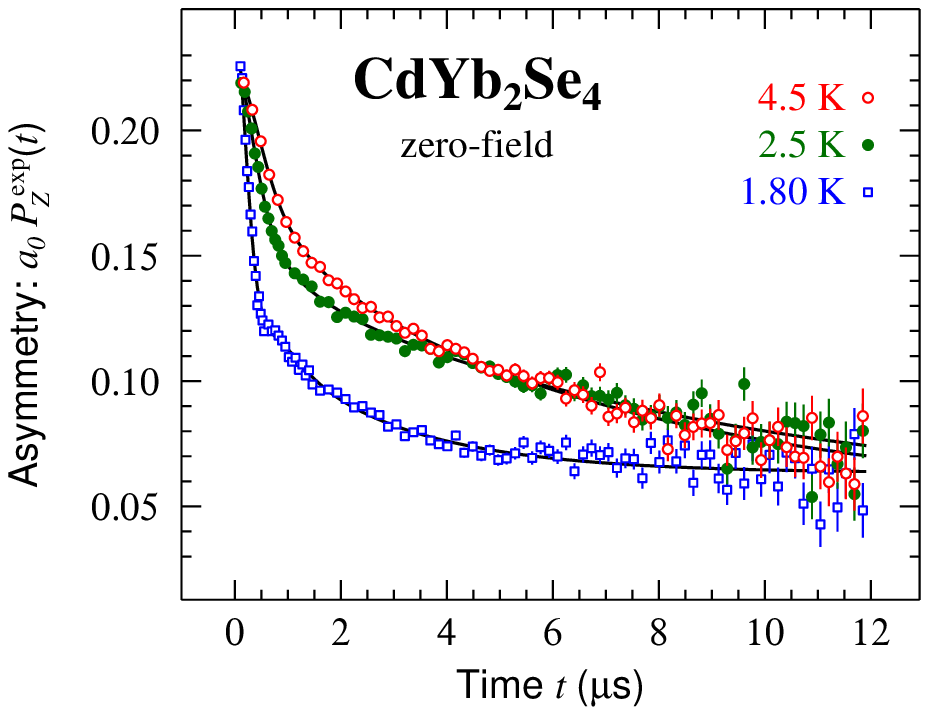}}
\put(55,310){(a)}
\put(16,0)
{\includegraphics[width=0.82\linewidth ]{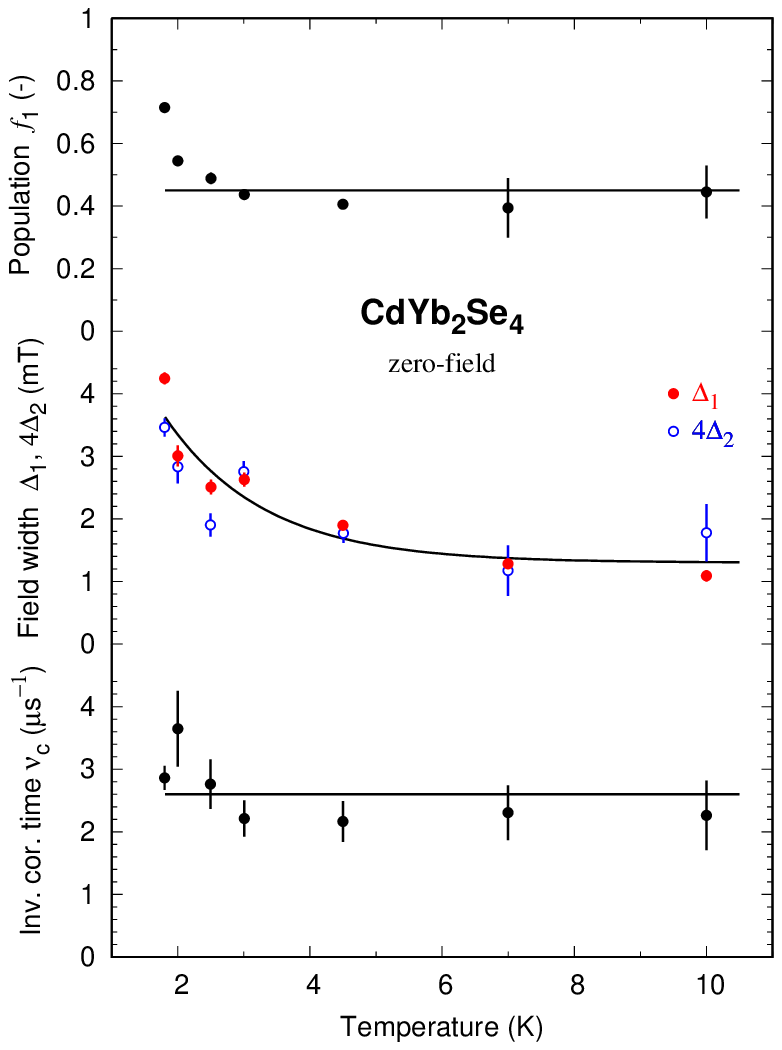}}
\put(55,30){(b)}
\end{picture}
\caption{(color online). 
(a) A selection of zero-field $\mu$SR spectra recorded in the correlated paramagnetic phase of CdYb$_2$Se$_4$. Solid lines result from fits as described in the main text. (b) Thermal dependence of the fitting parameters. The full lines are guides to the eye. 
}
\label{CdYb2Se4_muSR_ZF_parameters}
\end{figure}

\begin{figure}
\includegraphics[width=0.9\linewidth ]{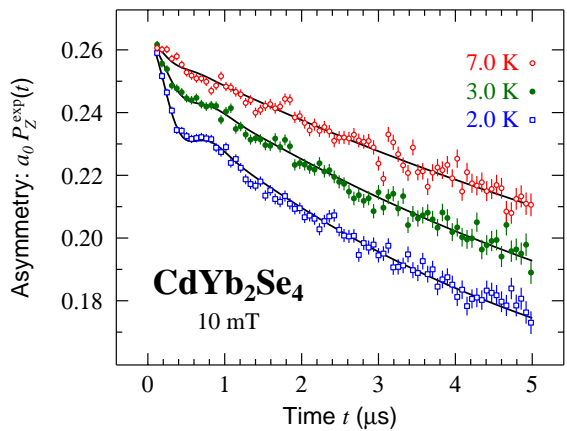}
\caption{(color online). 
$\mu$SR spectra recorded under a 10~mT longitudinal field in the paramagnetic phase of CdYb$_2$Se$_4$. Solid lines result from fits with the dynamical Kubo-Toyabe model. The pronounced shoulder at short times for the 2.0~K spectrum is a signature of the quasi-static dynamics of the magnetic field at the muon site.}
\label{CdYb2Se4_muSR_10mT}
\end{figure}

It has been recently shown that careful analyses of $\mu$SR spectra recorded in correlated paramagnetic regimes can yield useful information \cite{Maisuradze15,Dalmas17a}. The spectra for the two compounds of interest here are best described as the weighted sum of two dynamical Kubo-Toyabe functions; see Fig.~\ref{CdYb2Se4_muSR_ZF_parameters}(a) for an illustration in the case of CdYb$_2$Se$_4$.
Figure~\ref{CdYb2Se4_muSR_ZF_parameters}(b) displays the temperature dependence of the parameters extracted from a combined fit of the seven zero-field (ZF) spectra recorded for CdYb$_2$Se$_4$: the relative population of one of the two muon sites $f_1$, the field widths $\Delta_i$ and the common inverse correlation time $\nu_{\rm c}$. Similar spectra (not shown) were obtained for CdYb$_2$S$_4$. In fact, these results are reminiscent of those recently published for Yb$_2$Ti$_2$O$_7$, Yb$_2$Sn$_2$O$_7$, and Nd$_2$Sn$_2$O$_7$ \cite{Maisuradze15,Dalmas17a}: a temperature independent $\nu_{\rm c}$ in the range of the inverse microsecond and an increase in field width as the compound is cooled to $T_{\rm c}$.  At first sight, one could be surprised by the small $\nu_{\rm c}$ values. The strong effect of an external field as small as 10~mT on the spectra is an additional proof of the quasi-static spin dynamics at play; see Fig.~\ref{CdYb2Se4_muSR_10mT}. 
As expected, the $f_1$ parameter is essentially found to be temperature independent, with the exception of the region close to $T_{\rm c}$ where features associated with the magnetic transition could not be grasped by the model.

\section{Discussion}
\label{Discussion}

We first stress that the results from Ref.~\onlinecite{Higo17} and ours compare favorably when comparison is possible, i.e.\ whenever data for a given technique are available from both studies. This means that samples of different origin have similar responses.  The two compounds of the current study show a very similar behavior with a notable exception. Whereas the temperature dependence of the order parameter follows the prediction from a mean-field model based on the Brillouin function for CdYb$_2$S$_4$, this is not the case for the selenide (Figs.~\ref{CdYb2S4_moment} and \ref{CdYb2Se4_moment}). We find the magnetic moment and even more convincingly the spontaneous field at the muon site, which is expected to be proportional to it, to drop more slowly. This is quantitatively illustrated by the difference in the exponent $\beta$ resulting from the fits shown in the two figures. A possibility could be that the dimensionality of the magnetic interactions is reduced in the selenide compared to the sulfide. The difference must be subtle since it is not reflected in the macroscopic data reported in Sect.\ \ref{macroscopic}.

We now discuss the observed magnetic structures and compare them with structures already established for pyrochlore oxides. Then we shall consider the spin dynamics in the ordered state, followed by a discussion of the dynamics in the correlated paramagnetic regime.

Due to powder averaging, it is not possible from our measurements to determine which of the two basis vectors of $\Gamma_5$ describes the magnetic structures of the two investigated spinels \cite{Poole07} \footnote{Electron spin resonance measurements suggest the CdYb$_2$S$_4$ magnetic structure to be described by $\psi_2$ rather than $\psi_3$ \cite{Yoshizawa15}. Further neutron diffraction, e.g. in an applied magnetic field, would be welcome for a confirmation.}. However, we already know that we are dealing with antiferromagnetic compounds with magnetic moments perpendicular to the local three-fold axis. 
This is in clear contrast to the pyrochlore ytterbium titanate and stannate which are splayed ferromagnets \cite{note_Yb2Ti2O7_Yb2Sn2O7}. These results confirm that for a given rare-earth ion the spinel chalcogenides and pyrochlore oxides can display very different magnetic properties such as the magnetic structures. Based on the work of Ref.~\onlinecite{Lago10}, it is tempting to attribute the difference to the CEFs. However, this approach is clearly not comprehensive since, based on the sign of the second order Stevens parameter which is the same for the Er$^{3+}$ and Yb$^{3+}$ ions, one would expect CdYb$_2$Se$_4$ to have the same easy-axis anisotropy as CdEr$_2$Se$_4$ \cite{Lago10}, which is not observed. In fact the exchange interaction anisotropy could explain the difference, as for the Er$_2$Ti$_2$O$_7$ -- Yb$_2$Ti$_2$O$_7$ pair.

In Table~\ref{magnetic_structure_table}
\begin{table*}
\begin{center}
  \caption{Magnetic space groups associated with each mode of the different irreps for ${\bf k} = (000)$ structures in insulating pyrochlore magnets. The magnetic space groups have been determined using the Bilbao Crystallographic Server \cite{Perez15}. For reference, the crystallographic space group and the magnetic moment components at each rare earth site are indicated. The site coordinates correspond to neighboring rare earths forming the corners of a regular tetrahedron. A dash sign means that the cell content is identical to the cell above. The prime superscript in some magnetic space groups indicates that the symmetry operation is combined with time reversal.
Examples of nature realization of different magnetic structures are given in the last columns together with references. Er$_2$Ge$_2$O$_7$ \cite{Dun15} is suggested to order according to mode $\psi_3$ [note (i)], while Yb$_2$Ge$_2$O$_7$ orders within $\Gamma_5$ \cite{Hallas16a}, i.e.\ the mode combination is unknown. The magnetic structure of Gd$_2$Sn$_2$O$_7$ is described by one of the three $\Gamma_7$ modes.\footnote{Notice that neutron powder diffraction, at least in zero field, does not allow to distinguish modes such as $\psi_4$, $\psi_5$ and $\psi_6$. If the actual Gd$_2$Sn$_2$O$_7$ structure is a linear combination of two or three of the $\Gamma_7$ modes, the magnetic space group symmetry is lower than I4$^\prime_1$/amd$^\prime$.}  The two spinel compounds of the current study CdYb$_2$S$_4$ and CdYb$_2$Se$_4$ order according to the $\psi_2$ or $\psi_3$ mode ($\Gamma_5$) or a linear combination of them. In the latter case their magnetic space group would be Fddd, like their space group. Notice that the Yb$_2$Ti$_2$O$_7$ magnetic structure does not belong to any of the mentioned representations. Its symmetry is lower: the magnetic space group is Im$^\prime$m$^\prime$a and the corresponding structural space group is Imma \cite{note_Yb2Ti2O7_Yb2Sn2O7}. 
}
  \label{magnetic_structure_table}
  \begin{tabular}{l@{\hskip 3mm}l@{\hskip 3mm}ccc@{\hskip 3mm}ccc@{\hskip 3mm}ccc@{\hskip 3mm}ccc@{\hskip 3mm}l@{\hskip 3mm}l@{\hskip 3mm}cl}
\hline\hline
irrep      & mode     & \multicolumn{12}{c}{site} & magnetic & space & & Examples or note \cr
           &          & \multicolumn{12}{c}{}     & group    & group & & \cr\hline
           &          & \multicolumn{3}{c}{\hspace{-3mm}$\frac{1}{2}\frac{1}{2}\frac{1}{2}$} & \multicolumn{3}{c}{\hspace{-3mm}$\frac{1}{2}\frac{1}{4}\frac{1}{4}$} & \multicolumn{3}{c}{\hspace{-3mm}$\frac{1}{4}\frac{1}{2}\frac{1}{4}$} & \multicolumn{3}{c}{\hspace{-3mm}$\frac{1}{4}\frac{1}{4}\frac{1}{2}$} & & & & \cr
           &          & $m_x$ & $m_y$ & $m_z$ & $m_x$ & $m_y$ & $m_z$ &  $m_x$ & $m_y$ & $m_z$ & $m_x$ & $m_y$ & $m_z$ & & & & \cr \hline
$\Gamma_3$ & $\psi_1$ & 1 & 1 & 1 & 1 & $\overline{1}$ & $\overline{1}$ & $\overline{1}$ & 1 & $\overline{1}$ & $\overline{1}$ & $\overline{1}$ & 1 & Fd$\overline{3}$m$^\prime$ & Fd$\overline{3}$m & & Nd$_2$Sn$_2$O$_7$ \cite{Bertin15}, Nd$_2$Zr$_2$O$_7$ \cite{Lhotel15}, \cr
& & & & & & & & & & & & & & & & & and Nd$_2$Hf$_2$O$_7$ \cite{Anand15} \cr
$\Gamma_5$ & $\psi_2$ & 1 & 1 & $\overline{2}$ & 1 & $\overline{1}$ & 2 & $\overline{1}$ & 1 & 2 & $\overline{1}$ & $\overline{1}$ & $\overline{2}$ & I4$^\prime_1$/am$^\prime$d  & I4$_1$/amd & & Er$_2$Ti$_2$O$_7$ \cite{Champion03,Poole07} \cr
---       & $\psi_3$ & $\overline{1}$ & 1 & 0 & $\overline{1}$ & $\overline{1}$ & 0 & 1 & 1 & 0 & 1 & $\overline{1}$ & 0 & I4$_1$/amd                & --- & & (i) \cr
$\Gamma_7$ & $\psi_4$ & 1 & $\overline{1}$ & 0 & $\overline{1}$ & $\overline{1}$ & 0 & 1 & 1 & 0 & $\overline{1}$ & 1 & 0 & I4$^\prime_1$/amd$^\prime$ & --- & &  \cr
---       & $\psi_5$ & 0 & 1 & $\overline{1}$ & 0 & $\overline{1}$ & 1 & 0 & $\overline{1}$ & $\overline{1}$ & 0 & 1 & 1 & ---                      & --- & \cr
---       & $\psi_6$ & $\overline{1}$ & 0 & 1 & 1 & 0 & 1 & 1 & 0 & $\overline{1}$ & $\overline{1}$ & 0 & $\overline{1}$ & ---                      & --- & \cr
$\Gamma_9$ & $\psi_7$ & 1 & 1 & 0 & $\overline{1}$ & 1 & 0 & 1 & $\overline{1}$ & 0 & $\overline{1}$ & $\overline{1}$ & 0 & I4$_1$/am$^\prime$d$^\prime$ & --- & \multirow{2}{*}{$\left.\vphantom{\begin{tabular}{c}1\\1\end{tabular}}\right\}$} & Tb$_2$Sn$_2$O$_7$ \cite{Mirebeau05,Dalmas06} \cr
---       & $\psi_8$ & 0 & 0 & 1 & 0 & 0 & 1 & 0 & 0 & 1 & 0 & 0 & 1 & ---                    & --- & & and Yb$_2$Sn$_2$O$_7$ \cite{Yaouanc13,Lago14}\cr
---       & $\psi_9$ & 0 & 1 & 1 & 0 & $\overline{1}$ & $\overline{1}$ & 0 & $\overline{1}$ & 1 & 0 & 1 & $\overline{1}$ & ---                    & --- & & \cr
---       & $\psi_{10}$ & 1 & 0 & 0 & 1 & 0 & 0 & 1 & 0 & 0 & 1 & 0 & 0 & --- & --- & & \cr
---       & $\psi_{11}$ & 1 & 0 & 1 & $\overline{1}$ & 0 & 1 & $\overline{1}$ & 0 & $\overline{1}$ & 1 & 0 & $\overline{1}$ & --- & --- & & \cr
---       & $\psi_{12}$ & 0 & 1 & 0 & 0 & 1 & 0 & 0 & 1 & 0 & 0 & 1 & 0 & --- & --- & & \cr
\hline\hline
  \end{tabular}
\end{center}
\end{table*}
we provide the magnetic and structural space groups corresponding to the modes associated with the four $\Gamma_{3,5,7,9}$ irreps which are found for magnetic structures with a ${\bf k}= (000)$ propagation vector. In addition we list the related rare earth pyrochlores and spinel chalcogenides for which the magnetic structures have been determined, restricting ourselves to insulating systems where only the rare earth sublattice is magnetic \footnote{The pyrochlore Gd$_2$Ti$_2$O$_7$ is not listed because its propagation wavevector is ${\bf k}= (1/2,1/2,1/2)$. In fact, its magnetic structure is still under debate \cite{Paddison15}.}. Remarkably, only the neodymium oxides keep their paramagnetic cubic crystal structure as they order magnetically, thanks to the all-in--all-out magnetic structure they adopt. 
The diversity of magnetic structures illustrates the ground state delicate nature resulting from frustration. A consistent theoretical modeling of this large variety would be welcome. A first step theoretical analysis has just been published \cite{Yan17}. It takes for granted ${\bf k} = (000)$ and determines the magnetic ground state as a function of Curnoe's exchange parameters \cite{Curnoe08} using a set of complementary approximate methods. A justification of the ubiquitous ${\bf k} = (000)$ would be worthwhile. In addition, this approach is still only valid for fictitious spinor crystal-field ground states.

We now consider the $\mu$SR data in the ordered magnetic states. Spontaneous fields are observed for both compounds. However, scaling them with the spontaneous field and the magnetic moment of Cr reported for the chromium thio-spinel CdCr$_2$S$_4$ \cite{Hartmann13,Baltzer65}, we compute an ytterbium magnetic moment $\approx 0.1 \, \mu_{\rm B}$. This is too small by nearly an order of magnitude.
In fact, the absence of a spontaneous field in magnetically ordered pyrochlores is not a rare feature \cite{Dalmas16a}. It has been argued to be the signature of the dynamical nature of the magnetic ground state \cite{Dalmas06}. It could be related to the fragmentation of the Yb$^{3+}$ magnetic moments into two parts: static and dynamical components. The latter component could wash out the spontaneous field to a certain extent in the $\mu$SR time scale. It would be complete for Tb$_2$Sn$_2$O$_7$ for example, and partial for the compounds of interest in this paper. A picture within a classical physics framework has been presented \cite{Brooks14}. An extension incorporating quantum mechanics is urgently needed.

As usual for geometrically frustrated materials, persistent spin dynamics is present in the ordered state. This is detected through a measured temperature independent spin-lattice relaxation rate and has been taken as the signature of spin loops \cite{Yaouanc15}. This is again a proof of exotic spin fluctuations in the ordered state, albeit of origin different from the anomalously small spontaneous fields. In fact, persistent spin dynamics is ubiquitous no matter the ground state of the compound, i.e.\ magnetically ordered or not.

Finally, we turn to the correlated paramagnetic regime probed up to $\approx 10$~K. Similar results are again obtained for the two compounds. The most remarkable feature is the temperature independence deep in the paramagnetic phase, up to 10~K, of the inverse of the correlation time of the magnetic fluctuations with a value  $\approx 3 \, \mu{\rm s}^{-1}$. Evidence is now accumulating for spin dynamics in this time range or even slower \cite{Orendac16,Dalmas17a,Maisuradze15}. It could reflect a double spin-flip tunneling relaxation mechanism \cite{Bloembergen59} as recently argued \cite{Orendac16}. Interestingly, the approach to the magnetic phase transition is not seen through spin dynamics, but in the noticeable increase of the static field width as the compound is cooled. This is still to be understood.

\section{Conclusions}
\label{Conclusions}

An experimental study of two spinel ytterbium chalcogenides has been presented. In contrast to the splayed ferromagnetic order of the pyrochlore ytterbium stannate and titanate, the spinels are antiferromagnets with moments perpendicular to the local three-fold axis. Their magnetic structures can be described by the $\Gamma_5$ representation. Spontaneous magnetic fields at the muon site are observed, albeit with extremely reduced values. This suggests the magnetic (ordered) ground states to be of dynamical origin. Persistent spin dynamics observed through the spin lattice-relaxation channel are an additional proof of it. Quantum spin tunneling is detected in the correlated paramagnetic regime. 

These magnetic properties invite us to consider tetrahedra of spins rather than single-ion spins as the building blocks in the correlated paramagnetic regime and its extension in the ordered states \cite{Yaouanc15}. In fact it seems that the dynamical magnetic properties of the lattice of corner-sharing tetrahedra owe much to their topology since these properties are shared by a large number of compounds, no matter the nature of their magnetic structure. Interestingly, tetrahedra of spins have recently been considered as key ingredients for a theoretical modeling of the physics in the magnetically ordered states of corner-sharing tetrahedra systems \cite{Yan17}.

From the experimental viewpoint, a spectroscopic determination of the CEFs for the different compounds is needed to complete the experimental physical picture. Combining results from spinels with different rare-earth ions for that purpose would be of great help to pinpoint the six crystal-field parameters \cite{Bertin12,Ruminy16}.  

\begin{acknowledgments}
We are grateful to D. Ryan for discussions. This research project has been partially supported by the European Commission under the 7th Framework Programme through the `Research Infrastructures' action of the `Capacities' Programme, Contract No: CP-CSA\_INFRA-2008-1.1.1 Number 226507-NMI3. Part of this work was performed at the Institut Laue Langevin, Grenoble, France, the ISIS pulsed muon facility, STFC Rutherford Appleton Laboratory, Chilton, United Kingdom, and the Swiss muon source of the Paul Scherrer Institute, Villigen, Switzerland.
\end{acknowledgments}

\bibliography{reference,CdYb2X4_note}

\begin{thebibliography}{61}%
\makeatletter
\providecommand \@ifxundefined [1]{%
 \@ifx{#1\undefined}
}%
\providecommand \@ifnum [1]{%
 \ifnum #1\expandafter \@firstoftwo
 \else \expandafter \@secondoftwo
 \fi
}%
\providecommand \@ifx [1]{%
 \ifx #1\expandafter \@firstoftwo
 \else \expandafter \@secondoftwo
 \fi
}%
\providecommand \natexlab [1]{#1}%
\providecommand \enquote  [1]{``#1''}%
\providecommand \bibnamefont  [1]{#1}%
\providecommand \bibfnamefont [1]{#1}%
\providecommand \citenamefont [1]{#1}%
\providecommand \href@noop [0]{\@secondoftwo}%
\providecommand \href [0]{\begingroup \@sanitize@url \@href}%
\providecommand \@href[1]{\@@startlink{#1}\@@href}%
\providecommand \@@href[1]{\endgroup#1\@@endlink}%
\providecommand \@sanitize@url [0]{\catcode `\\12\catcode `\$12\catcode
  `\&12\catcode `\#12\catcode `\^12\catcode `\_12\catcode `\%12\relax}%
\providecommand \@@startlink[1]{}%
\providecommand \@@endlink[0]{}%
\providecommand \url  [0]{\begingroup\@sanitize@url \@url }%
\providecommand \@url [1]{\endgroup\@href {#1}{\urlprefix }}%
\providecommand \urlprefix  [0]{URL }%
\providecommand \Eprint [0]{\href }%
\providecommand \doibase [0]{http://dx.doi.org/}%
\providecommand \selectlanguage [0]{\@gobble}%
\providecommand \bibinfo  [0]{\@secondoftwo}%
\providecommand \bibfield  [0]{\@secondoftwo}%
\providecommand \translation [1]{[#1]}%
\providecommand \BibitemOpen [0]{}%
\providecommand \bibitemStop [0]{}%
\providecommand \bibitemNoStop [0]{.\EOS\space}%
\providecommand \EOS [0]{\spacefactor3000\relax}%
\providecommand \BibitemShut  [1]{\csname bibitem#1\endcsname}%
\let\auto@bib@innerbib\@empty
\bibitem [{\citenamefont {Gardner}\ \emph {et~al.}(2010)\citenamefont
  {Gardner}, \citenamefont {Gingras},\ and\ \citenamefont
  {Greedan}}]{Gardner10}%
  \BibitemOpen
  \bibfield  {author} {\bibinfo {author} {\bibfnamefont {Jason~S.}\
  \bibnamefont {Gardner}}, \bibinfo {author} {\bibfnamefont {Michel J.~P.}\
  \bibnamefont {Gingras}}, \ and\ \bibinfo {author} {\bibfnamefont {John~E.}\
  \bibnamefont {Greedan}},\ }\bibfield  {title} {\enquote {\bibinfo {title}
  {Magnetic pyrochlore oxides},}\ }\href {\doibase 10.1103/RevModPhys.82.53}
  {\bibfield  {journal} {\bibinfo  {journal} {Rev. Mod. Phys.}\ }\textbf
  {\bibinfo {volume} {82}},\ \bibinfo {pages} {53} (\bibinfo {year}
  {2010})}\BibitemShut {NoStop}%
\bibitem [{\citenamefont {Gingras}\ and\ \citenamefont
  {McClarty}(2014)}]{Gingras14}%
  \BibitemOpen
  \bibfield  {author} {\bibinfo {author} {\bibfnamefont {M.~J.~P.}\
  \bibnamefont {Gingras}}\ and\ \bibinfo {author} {\bibfnamefont {P.~A.}\
  \bibnamefont {McClarty}},\ }\bibfield  {title} {\enquote {\bibinfo {title}
  {Quantum spin ice: a search for gapless quantum spin liquids in pyrochlore
  magnets},}\ }\href {\doibase 10.1088/0034-4885/77/5/056501} {\bibfield
  {journal} {\bibinfo  {journal} {Rep. Prog. Phys.}\ }\textbf {\bibinfo
  {volume} {77}},\ \bibinfo {pages} {056501} (\bibinfo {year}
  {2014})}\BibitemShut {NoStop}%
\bibitem [{\citenamefont {Lhotel}\ \emph {et~al.}(2015)\citenamefont {Lhotel},
  \citenamefont {Petit}, \citenamefont {Guitteny}, \citenamefont {Florea},
  \citenamefont {Ciomaga~Hatnean}, \citenamefont {Colin}, \citenamefont
  {Ressouche}, \citenamefont {Lees},\ and\ \citenamefont
  {Balakrishnan}}]{Lhotel15}%
  \BibitemOpen
  \bibfield  {author} {\bibinfo {author} {\bibfnamefont {E.}~\bibnamefont
  {Lhotel}}, \bibinfo {author} {\bibfnamefont {S.}~\bibnamefont {Petit}},
  \bibinfo {author} {\bibfnamefont {S.}~\bibnamefont {Guitteny}}, \bibinfo
  {author} {\bibfnamefont {O.}~\bibnamefont {Florea}}, \bibinfo {author}
  {\bibfnamefont {M.}~\bibnamefont {Ciomaga~Hatnean}}, \bibinfo {author}
  {\bibfnamefont {C.}~\bibnamefont {Colin}}, \bibinfo {author} {\bibfnamefont
  {E.}~\bibnamefont {Ressouche}}, \bibinfo {author} {\bibfnamefont {M.~R.}\
  \bibnamefont {Lees}}, \ and\ \bibinfo {author} {\bibfnamefont
  {G.}~\bibnamefont {Balakrishnan}},\ }\bibfield  {title} {\enquote {\bibinfo
  {title} {Fluctuations and all-in--all-out ordering in dipole-octupole
  $\mathrm{Nd}_{2}\mathrm{Zr}_{2}\mathrm{O}_{7}$},}\ }\href {\doibase
  10.1103/PhysRevLett.115.197202} {\bibfield  {journal} {\bibinfo  {journal}
  {Phys. Rev. Lett.}\ }\textbf {\bibinfo {volume} {115}},\ \bibinfo {pages}
  {197202} (\bibinfo {year} {2015})}\BibitemShut {NoStop}%
\bibitem [{\citenamefont {Xu}\ \emph {et~al.}(2015)\citenamefont {Xu},
  \citenamefont {Anand}, \citenamefont {Bera}, \citenamefont {Frontzek},
  \citenamefont {Abernathy}, \citenamefont {Casati}, \citenamefont
  {Siemensmeyer},\ and\ \citenamefont {Lake}}]{Xu15}%
  \BibitemOpen
  \bibfield  {author} {\bibinfo {author} {\bibfnamefont {J.}~\bibnamefont
  {Xu}}, \bibinfo {author} {\bibfnamefont {V.~K.}\ \bibnamefont {Anand}},
  \bibinfo {author} {\bibfnamefont {A.~K.}\ \bibnamefont {Bera}}, \bibinfo
  {author} {\bibfnamefont {M.}~\bibnamefont {Frontzek}}, \bibinfo {author}
  {\bibfnamefont {D.~L.}\ \bibnamefont {Abernathy}}, \bibinfo {author}
  {\bibfnamefont {N.}~\bibnamefont {Casati}}, \bibinfo {author} {\bibfnamefont
  {K.}~\bibnamefont {Siemensmeyer}}, \ and\ \bibinfo {author} {\bibfnamefont
  {B.}~\bibnamefont {Lake}},\ }\bibfield  {title} {\enquote {\bibinfo {title}
  {Magnetic structure and crystal-field states of the pyrochlore
  antiferromagnet \uppercase{N}d$_2$\uppercase{Z}r$_2$\uppercase{O}$_2$},}\
  }\href {\doibase 10.1103/PhysRevB.92.224430} {\bibfield  {journal} {\bibinfo
  {journal} {Phys. Rev. B}\ }\textbf {\bibinfo {volume} {92}},\ \bibinfo
  {pages} {224430} (\bibinfo {year} {2015})}\BibitemShut {NoStop}%
\bibitem [{\citenamefont {Xu}\ \emph {et~al.}(2016)\citenamefont {Xu},
  \citenamefont {Balz}, \citenamefont {Baines}, \citenamefont {Luetkens},\ and\
  \citenamefont {Lake}}]{Xu16}%
  \BibitemOpen
  \bibfield  {author} {\bibinfo {author} {\bibfnamefont {J.}~\bibnamefont
  {Xu}}, \bibinfo {author} {\bibfnamefont {C.}~\bibnamefont {Balz}}, \bibinfo
  {author} {\bibfnamefont {C.}~\bibnamefont {Baines}}, \bibinfo {author}
  {\bibfnamefont {H.}~\bibnamefont {Luetkens}}, \ and\ \bibinfo {author}
  {\bibfnamefont {B.}~\bibnamefont {Lake}},\ }\bibfield  {title} {\enquote
  {\bibinfo {title} {Spin dynamics of the ordered dipolar-octupolar
  pseudospin-$\frac{1}{2}$ pyrochlore
  \uppercase{N}d$_2$\uppercase{Z}r$_2$\uppercase{O}$_7$ probed by muon spin
  relaxation},}\ }\href {\doibase 10.1103/PhysRevB.94.064425} {\bibfield
  {journal} {\bibinfo  {journal} {Phys. Rev. B}\ }\textbf {\bibinfo {volume}
  {94}},\ \bibinfo {pages} {064425} (\bibinfo {year} {2016})}\BibitemShut
  {NoStop}%
\bibitem [{\citenamefont {Bonville}\ \emph {et~al.}(2016)\citenamefont
  {Bonville}, \citenamefont {Guitteny}, \citenamefont {Gukasov}, \citenamefont
  {Mirebeau}, \citenamefont {Petit}, \citenamefont {Decorse}, \citenamefont
  {Hatnean~Ciomaga},\ and\ \citenamefont {Balakrishnan}}]{Bonville16}%
  \BibitemOpen
  \bibfield  {author} {\bibinfo {author} {\bibfnamefont {P.}~\bibnamefont
  {Bonville}}, \bibinfo {author} {\bibfnamefont {S.}~\bibnamefont {Guitteny}},
  \bibinfo {author} {\bibfnamefont {A.}~\bibnamefont {Gukasov}}, \bibinfo
  {author} {\bibfnamefont {I.}~\bibnamefont {Mirebeau}}, \bibinfo {author}
  {\bibfnamefont {S.}~\bibnamefont {Petit}}, \bibinfo {author} {\bibfnamefont
  {C.}~\bibnamefont {Decorse}}, \bibinfo {author} {\bibfnamefont
  {M.}~\bibnamefont {Hatnean~Ciomaga}}, \ and\ \bibinfo {author} {\bibfnamefont
  {G.}~\bibnamefont {Balakrishnan}},\ }\bibfield  {title} {\enquote {\bibinfo
  {title} {Magnetic properties and crystal field in
  $\mathrm{Pr}_{2}\mathrm{Zr}_{2}\mathrm{O}_{7}$},}\ }\href {\doibase
  10.1103/PhysRevB.94.134428} {\bibfield  {journal} {\bibinfo  {journal} {Phys.
  Rev. B}\ }\textbf {\bibinfo {volume} {94}},\ \bibinfo {pages} {134428}
  (\bibinfo {year} {2016})}\BibitemShut {NoStop}%
\bibitem [{\citenamefont {Petit}\ \emph {et~al.}()\citenamefont {Petit},
  \citenamefont {Lhotel}, \citenamefont {Canals}, \citenamefont
  {Ciomaga~Hatnean}, \citenamefont {Ollivier}, \citenamefont {Mutka},
  \citenamefont {Ressouche}, \citenamefont {Wildes}, \citenamefont {Lees},\
  and\ \citenamefont {Balakrishnan}}]{Petit16}%
  \BibitemOpen
  \bibfield  {author} {\bibinfo {author} {\bibfnamefont {S.}~\bibnamefont
  {Petit}}, \bibinfo {author} {\bibfnamefont {E.}~\bibnamefont {Lhotel}},
  \bibinfo {author} {\bibfnamefont {B.}~\bibnamefont {Canals}}, \bibinfo
  {author} {\bibfnamefont {M.}~\bibnamefont {Ciomaga~Hatnean}}, \bibinfo
  {author} {\bibfnamefont {J.}~\bibnamefont {Ollivier}}, \bibinfo {author}
  {\bibfnamefont {H.}~\bibnamefont {Mutka}}, \bibinfo {author} {\bibfnamefont
  {E.}~\bibnamefont {Ressouche}}, \bibinfo {author} {\bibfnamefont {A.~R.}\
  \bibnamefont {Wildes}}, \bibinfo {author} {\bibfnamefont {M.~R.}\
  \bibnamefont {Lees}}, \ and\ \bibinfo {author} {\bibfnamefont
  {G.}~\bibnamefont {Balakrishnan}},\ }\bibfield  {title} {\enquote {\bibinfo
  {title} {Observation of magnetic fragmentation in spin ice},}\ }\href
  {\doibase 10.1038/nphys3710} {\bibfield  {journal} {\bibinfo  {journal} {Nat.
  Phys.}\ }\textbf {\bibinfo {volume} {12}},\ \bibinfo {pages}
  {746}}\BibitemShut {NoStop}%
\bibitem [{\citenamefont {Wen}\ \emph {et~al.}(2017)\citenamefont {Wen},
  \citenamefont {Koohpayeh}, \citenamefont {Ross}, \citenamefont {Trump},
  \citenamefont {McQueen}, \citenamefont {Kimura}, \citenamefont {Nakatsuji},
  \citenamefont {Qiu}, \citenamefont {Pajerowski}, \citenamefont {Copley},\
  and\ \citenamefont {Broholm}}]{Wen17}%
  \BibitemOpen
  \bibfield  {author} {\bibinfo {author} {\bibfnamefont {J.-J.}\ \bibnamefont
  {Wen}}, \bibinfo {author} {\bibfnamefont {S.~M.}\ \bibnamefont {Koohpayeh}},
  \bibinfo {author} {\bibfnamefont {K.~A.}\ \bibnamefont {Ross}}, \bibinfo
  {author} {\bibfnamefont {B.~A.}\ \bibnamefont {Trump}}, \bibinfo {author}
  {\bibfnamefont {T.~M.}\ \bibnamefont {McQueen}}, \bibinfo {author}
  {\bibfnamefont {K.}~\bibnamefont {Kimura}}, \bibinfo {author} {\bibfnamefont
  {S.}~\bibnamefont {Nakatsuji}}, \bibinfo {author} {\bibfnamefont
  {Y.}~\bibnamefont {Qiu}}, \bibinfo {author} {\bibfnamefont {D.~M.}\
  \bibnamefont {Pajerowski}}, \bibinfo {author} {\bibfnamefont {J.~R.~D.}\
  \bibnamefont {Copley}}, \ and\ \bibinfo {author} {\bibfnamefont {C.~L.}\
  \bibnamefont {Broholm}},\ }\bibfield  {title} {\enquote {\bibinfo {title}
  {Disordered route to the coulomb quantum spin liquid: Random transverse
  fields on spin ice in $\mathrm{Pr}_{2}\mathrm{Zr}_{2}\mathrm{O}_{7}$},}\
  }\href {\doibase 10.1103/PhysRevLett.118.107206} {\bibfield  {journal}
  {\bibinfo  {journal} {Phys. Rev. Lett.}\ }\textbf {\bibinfo {volume} {118}},\
  \bibinfo {pages} {107206} (\bibinfo {year} {2017})}\BibitemShut {NoStop}%
\bibitem [{\citenamefont {Sibille}\ \emph {et~al.}(2016)\citenamefont
  {Sibille}, \citenamefont {Lhotel}, \citenamefont {Hatnean}, \citenamefont
  {Balakrishnan}, \citenamefont {F\aa{}k}, \citenamefont {Gauthier},
  \citenamefont {Fennell},\ and\ \citenamefont {Kenzelmann}}]{Sibille16}%
  \BibitemOpen
  \bibfield  {author} {\bibinfo {author} {\bibfnamefont {Romain}\ \bibnamefont
  {Sibille}}, \bibinfo {author} {\bibfnamefont {Elsa}\ \bibnamefont {Lhotel}},
  \bibinfo {author} {\bibfnamefont {Monica~Ciomaga}\ \bibnamefont {Hatnean}},
  \bibinfo {author} {\bibfnamefont {Geetha}\ \bibnamefont {Balakrishnan}},
  \bibinfo {author} {\bibfnamefont {Bj\"orn}\ \bibnamefont {F\aa{}k}}, \bibinfo
  {author} {\bibfnamefont {Nicolas}\ \bibnamefont {Gauthier}}, \bibinfo
  {author} {\bibfnamefont {Tom}\ \bibnamefont {Fennell}}, \ and\ \bibinfo
  {author} {\bibfnamefont {Michel}\ \bibnamefont {Kenzelmann}},\ }\bibfield
  {title} {\enquote {\bibinfo {title} {Candidate quantum spin ice in the
  pyrochlore $\mathrm{Pr}_{2}\mathrm{Hf}_{2}\mathrm{O}_{7}$},}\ }\href
  {\doibase 10.1103/PhysRevB.94.024436} {\bibfield  {journal} {\bibinfo
  {journal} {Phys. Rev. B}\ }\textbf {\bibinfo {volume} {94}},\ \bibinfo
  {pages} {024436} (\bibinfo {year} {2016})}\BibitemShut {NoStop}%
\bibitem [{\citenamefont {Anand}\ \emph {et~al.}(2015)\citenamefont {Anand},
  \citenamefont {Bera}, \citenamefont {Xu}, \citenamefont {Herrmannsd\"orfer},
  \citenamefont {Ritter},\ and\ \citenamefont {Lake}}]{Anand15}%
  \BibitemOpen
  \bibfield  {author} {\bibinfo {author} {\bibfnamefont {V.~K.}\ \bibnamefont
  {Anand}}, \bibinfo {author} {\bibfnamefont {A.~K.}\ \bibnamefont {Bera}},
  \bibinfo {author} {\bibfnamefont {J.}~\bibnamefont {Xu}}, \bibinfo {author}
  {\bibfnamefont {T.}~\bibnamefont {Herrmannsd\"orfer}}, \bibinfo {author}
  {\bibfnamefont {C.}~\bibnamefont {Ritter}}, \ and\ \bibinfo {author}
  {\bibfnamefont {B.}~\bibnamefont {Lake}},\ }\bibfield  {title} {\enquote
  {\bibinfo {title} {Observation of long-range magnetic ordering in pyrohafnate
  $\mathrm{Nd}_{2}\mathrm{Hf}_{2}\mathrm{O}_{7}$: A neutron diffraction
  study},}\ }\href {\doibase 10.1103/PhysRevB.92.184418} {\bibfield  {journal}
  {\bibinfo  {journal} {Phys. Rev. B}\ }\textbf {\bibinfo {volume} {92}},\
  \bibinfo {pages} {184418} (\bibinfo {year} {2015})}\BibitemShut {NoStop}%
\bibitem [{\citenamefont {Anand}\ \emph {et~al.}(2016)\citenamefont {Anand},
  \citenamefont {Opherden}, \citenamefont {Xu}, \citenamefont {Adroja},
  \citenamefont {Islam}, \citenamefont {Herrmannsd\"orfer}, \citenamefont
  {Hornung}, \citenamefont {Sch\"onemann}, \citenamefont {Uhlarz},
  \citenamefont {Walker}, \citenamefont {Casati},\ and\ \citenamefont
  {Lake}}]{Anand16}%
  \BibitemOpen
  \bibfield  {author} {\bibinfo {author} {\bibfnamefont {V.~K.}\ \bibnamefont
  {Anand}}, \bibinfo {author} {\bibfnamefont {L.}~\bibnamefont {Opherden}},
  \bibinfo {author} {\bibfnamefont {J.}~\bibnamefont {Xu}}, \bibinfo {author}
  {\bibfnamefont {D.~T.}\ \bibnamefont {Adroja}}, \bibinfo {author}
  {\bibfnamefont {A.~T. M.~N.}\ \bibnamefont {Islam}}, \bibinfo {author}
  {\bibfnamefont {T.}~\bibnamefont {Herrmannsd\"orfer}}, \bibinfo {author}
  {\bibfnamefont {J.}~\bibnamefont {Hornung}}, \bibinfo {author} {\bibfnamefont
  {R.}~\bibnamefont {Sch\"onemann}}, \bibinfo {author} {\bibfnamefont
  {M.}~\bibnamefont {Uhlarz}}, \bibinfo {author} {\bibfnamefont {H.~C.}\
  \bibnamefont {Walker}}, \bibinfo {author} {\bibfnamefont {N.}~\bibnamefont
  {Casati}}, \ and\ \bibinfo {author} {\bibfnamefont {B.}~\bibnamefont
  {Lake}},\ }\bibfield  {title} {\enquote {\bibinfo {title} {Physical
  properties of the candidate quantum spin-ice system
  $\mathrm{Pr}_{2}\mathrm{Hf}_{2}\mathrm{O}_{7}$},}\ }\href {\doibase
  10.1103/PhysRevB.94.144415} {\bibfield  {journal} {\bibinfo  {journal} {Phys.
  Rev. B}\ }\textbf {\bibinfo {volume} {94}},\ \bibinfo {pages} {144415}
  (\bibinfo {year} {2016})}\BibitemShut {NoStop}%
\bibitem [{\citenamefont {Wiebe}\ and\ \citenamefont {Hallas}(2015)}]{Wiebe15}%
  \BibitemOpen
  \bibfield  {author} {\bibinfo {author} {\bibfnamefont {C.~R.}\ \bibnamefont
  {Wiebe}}\ and\ \bibinfo {author} {\bibfnamefont {A.~M.}\ \bibnamefont
  {Hallas}},\ }\bibfield  {title} {\enquote {\bibinfo {title} {Frustration
  under pressure: Exotic magnetism in new pyrochlore oxides},}\ }\href
  {\doibase 10.1063/1.4916020} {\bibfield  {journal} {\bibinfo  {journal} {APL
  Materials}\ }\textbf {\bibinfo {volume} {3}},\ \bibinfo {pages} {041519}
  (\bibinfo {year} {2015})}\BibitemShut {NoStop}%
\bibitem [{\citenamefont {Hallas}\ \emph {et~al.}(2015)\citenamefont {Hallas},
  \citenamefont {Arevalo-Lopez}, \citenamefont {Sharma}, \citenamefont
  {Munsie}, \citenamefont {Attfield}, \citenamefont {Wiebe},\ and\
  \citenamefont {Luke}}]{Hallas15}%
  \BibitemOpen
  \bibfield  {author} {\bibinfo {author} {\bibfnamefont {A.~M.}\ \bibnamefont
  {Hallas}}, \bibinfo {author} {\bibfnamefont {A.~M.}\ \bibnamefont
  {Arevalo-Lopez}}, \bibinfo {author} {\bibfnamefont {A.~Z.}\ \bibnamefont
  {Sharma}}, \bibinfo {author} {\bibfnamefont {T.}~\bibnamefont {Munsie}},
  \bibinfo {author} {\bibfnamefont {J.~P.}\ \bibnamefont {Attfield}}, \bibinfo
  {author} {\bibfnamefont {C.~R.}\ \bibnamefont {Wiebe}}, \ and\ \bibinfo
  {author} {\bibfnamefont {G.~M.}\ \bibnamefont {Luke}},\ }\bibfield  {title}
  {\enquote {\bibinfo {title} {Magnetic frustration in lead pyrochlores},}\
  }\href {\doibase 10.1103/PhysRevB.91.104417} {\bibfield  {journal} {\bibinfo
  {journal} {Phys. Rev. B}\ }\textbf {\bibinfo {volume} {91}},\ \bibinfo
  {pages} {104417} (\bibinfo {year} {2015})}\BibitemShut {NoStop}%
\bibitem [{\citenamefont {Cai}\ \emph {et~al.}(2016)\citenamefont {Cai},
  \citenamefont {Cui}, \citenamefont {Li}, \citenamefont {Dun}, \citenamefont
  {Ma}, \citenamefont {dela Cruz}, \citenamefont {Jiao}, \citenamefont {Liao},
  \citenamefont {Sun}, \citenamefont {Li}, \citenamefont {Zhou}, \citenamefont
  {Goodenough}, \citenamefont {Zhou},\ and\ \citenamefont {Cheng}}]{Cai16}%
  \BibitemOpen
  \bibfield  {author} {\bibinfo {author} {\bibfnamefont {Y.~Q.}\ \bibnamefont
  {Cai}}, \bibinfo {author} {\bibfnamefont {Q.}~\bibnamefont {Cui}}, \bibinfo
  {author} {\bibfnamefont {X.}~\bibnamefont {Li}}, \bibinfo {author}
  {\bibfnamefont {Z.~L.}\ \bibnamefont {Dun}}, \bibinfo {author} {\bibfnamefont
  {J.}~\bibnamefont {Ma}}, \bibinfo {author} {\bibfnamefont {C.}~\bibnamefont
  {dela Cruz}}, \bibinfo {author} {\bibfnamefont {Y.~Y.}\ \bibnamefont {Jiao}},
  \bibinfo {author} {\bibfnamefont {J.}~\bibnamefont {Liao}}, \bibinfo {author}
  {\bibfnamefont {P.~J.}\ \bibnamefont {Sun}}, \bibinfo {author} {\bibfnamefont
  {Y.~Q.}\ \bibnamefont {Li}}, \bibinfo {author} {\bibfnamefont {J.~S.}\
  \bibnamefont {Zhou}}, \bibinfo {author} {\bibfnamefont {J.~B.}\ \bibnamefont
  {Goodenough}}, \bibinfo {author} {\bibfnamefont {H.~D.}\ \bibnamefont
  {Zhou}}, \ and\ \bibinfo {author} {\bibfnamefont {J.-G.}\ \bibnamefont
  {Cheng}},\ }\bibfield  {title} {\enquote {\bibinfo {title} {High-pressure
  synthesis and characterization of the effective pseudospin
  $\uppercase{S}=1/2$ \uppercase{XY} pyrochlores
  $\uppercase{R}_{2}$\uppercase{P}t$_{2}$\uppercase{O}$_{7}$ ($\uppercase{R}$ =
  \uppercase{E}r, \uppercase{Y}b)},}\ }\href {\doibase
  10.1103/PhysRevB.93.014443} {\bibfield  {journal} {\bibinfo  {journal} {Phys.
  Rev. B}\ }\textbf {\bibinfo {volume} {93}},\ \bibinfo {pages} {014443}
  (\bibinfo {year} {2016})}\BibitemShut {NoStop}%
\bibitem [{\citenamefont {Hallas}\ \emph
  {et~al.}(2016{\natexlab{a}})\citenamefont {Hallas}, \citenamefont {Sharma},
  \citenamefont {Cai}, \citenamefont {Munsie}, \citenamefont {Wilson},
  \citenamefont {Tachibana}, \citenamefont {Wiebe},\ and\ \citenamefont
  {Luke}}]{Hallas16b}%
  \BibitemOpen
  \bibfield  {author} {\bibinfo {author} {\bibfnamefont {A.~M.}\ \bibnamefont
  {Hallas}}, \bibinfo {author} {\bibfnamefont {A.~Z.}\ \bibnamefont {Sharma}},
  \bibinfo {author} {\bibfnamefont {Y.}~\bibnamefont {Cai}}, \bibinfo {author}
  {\bibfnamefont {T.~J.}\ \bibnamefont {Munsie}}, \bibinfo {author}
  {\bibfnamefont {M.~N.}\ \bibnamefont {Wilson}}, \bibinfo {author}
  {\bibfnamefont {M.}~\bibnamefont {Tachibana}}, \bibinfo {author}
  {\bibfnamefont {C.~R.}\ \bibnamefont {Wiebe}}, \ and\ \bibinfo {author}
  {\bibfnamefont {G.~M.}\ \bibnamefont {Luke}},\ }\bibfield  {title} {\enquote
  {\bibinfo {title} {Relief of frustration in the \uppercase{H}eisenberg
  pyrochlore antiferromagnet $\mathrm{Gd}_{2}\mathrm{Pt}_{2}\mathrm{O}_{7}$},}\
  }\href {\doibase 10.1103/PhysRevB.94.134417} {\bibfield  {journal} {\bibinfo
  {journal} {Phys. Rev. B}\ }\textbf {\bibinfo {volume} {94}},\ \bibinfo
  {pages} {134417} (\bibinfo {year} {2016}{\natexlab{a}})}\BibitemShut
  {NoStop}%
\bibitem [{\citenamefont {Li}\ \emph {et~al.}(2016)\citenamefont {Li},
  \citenamefont {Cai}, \citenamefont {Cui}, \citenamefont {Lin}, \citenamefont
  {Dun}, \citenamefont {Matsubayashi}, \citenamefont {Uwatoko}, \citenamefont
  {Sato}, \citenamefont {Kawae}, \citenamefont {Lv}, \citenamefont {Jin},
  \citenamefont {Zhou}, \citenamefont {Goodenough}, \citenamefont {Zhou},\ and\
  \citenamefont {Cheng}}]{Li16}%
  \BibitemOpen
  \bibfield  {author} {\bibinfo {author} {\bibfnamefont {X.}~\bibnamefont
  {Li}}, \bibinfo {author} {\bibfnamefont {Y.~Q.}\ \bibnamefont {Cai}},
  \bibinfo {author} {\bibfnamefont {Q.}~\bibnamefont {Cui}}, \bibinfo {author}
  {\bibfnamefont {C.~J.}\ \bibnamefont {Lin}}, \bibinfo {author} {\bibfnamefont
  {Z.~L.}\ \bibnamefont {Dun}}, \bibinfo {author} {\bibfnamefont
  {K.}~\bibnamefont {Matsubayashi}}, \bibinfo {author} {\bibfnamefont
  {Y.}~\bibnamefont {Uwatoko}}, \bibinfo {author} {\bibfnamefont
  {Y.}~\bibnamefont {Sato}}, \bibinfo {author} {\bibfnamefont {T.}~\bibnamefont
  {Kawae}}, \bibinfo {author} {\bibfnamefont {S.~J.}\ \bibnamefont {Lv}},
  \bibinfo {author} {\bibfnamefont {C.~Q.}\ \bibnamefont {Jin}}, \bibinfo
  {author} {\bibfnamefont {J.-S.}\ \bibnamefont {Zhou}}, \bibinfo {author}
  {\bibfnamefont {J.~B.}\ \bibnamefont {Goodenough}}, \bibinfo {author}
  {\bibfnamefont {H.~D.}\ \bibnamefont {Zhou}}, \ and\ \bibinfo {author}
  {\bibfnamefont {J.-G.}\ \bibnamefont {Cheng}},\ }\bibfield  {title} {\enquote
  {\bibinfo {title} {Long-range magnetic order in the \uppercase{H}eisenberg
  pyrochlore antiferromagnets
  $\mathrm{G}\mathrm{d}_{2}\mathrm{G}\mathrm{e}_{2}\mathrm{O}_{7}$ and
  $\mathrm{G}\mathrm{d}_{2}\mathrm{P}\mathrm{t}_{2}\mathrm{O}_{7}$ synthesized
  under high pressure},}\ }\href {\doibase 10.1103/PhysRevB.94.214429}
  {\bibfield  {journal} {\bibinfo  {journal} {Phys. Rev. B}\ }\textbf {\bibinfo
  {volume} {94}},\ \bibinfo {pages} {214429} (\bibinfo {year}
  {2016})}\BibitemShut {NoStop}%
\bibitem [{\citenamefont {Dun}\ \emph {et~al.}(2014)\citenamefont {Dun},
  \citenamefont {Lee}, \citenamefont {Choi}, \citenamefont {Hallas},
  \citenamefont {Wiebe}, \citenamefont {Gardner}, \citenamefont {Arrighi},
  \citenamefont {Freitas}, \citenamefont {Arevalo-Lopez}, \citenamefont
  {Attfield}, \citenamefont {Zhou},\ and\ \citenamefont {Cheng}}]{Dun14}%
  \BibitemOpen
  \bibfield  {author} {\bibinfo {author} {\bibfnamefont {Z.~L.}\ \bibnamefont
  {Dun}}, \bibinfo {author} {\bibfnamefont {M.}~\bibnamefont {Lee}}, \bibinfo
  {author} {\bibfnamefont {E.~S.}\ \bibnamefont {Choi}}, \bibinfo {author}
  {\bibfnamefont {A.~M.}\ \bibnamefont {Hallas}}, \bibinfo {author}
  {\bibfnamefont {C.~R.}\ \bibnamefont {Wiebe}}, \bibinfo {author}
  {\bibfnamefont {J.~S.}\ \bibnamefont {Gardner}}, \bibinfo {author}
  {\bibfnamefont {E.}~\bibnamefont {Arrighi}}, \bibinfo {author} {\bibfnamefont
  {R.~S.}\ \bibnamefont {Freitas}}, \bibinfo {author} {\bibfnamefont {A.~M.}\
  \bibnamefont {Arevalo-Lopez}}, \bibinfo {author} {\bibfnamefont {J.~P.}\
  \bibnamefont {Attfield}}, \bibinfo {author} {\bibfnamefont {H.~D.}\
  \bibnamefont {Zhou}}, \ and\ \bibinfo {author} {\bibfnamefont {J.~G.}\
  \bibnamefont {Cheng}},\ }\bibfield  {title} {\enquote {\bibinfo {title}
  {Chemical pressure effects on magnetism in the quantum spin liquid candidates
  \uppercase{Y}b${}_{2}{\uppercase{x}}_{2}$\uppercase{O}${}_{7}$
  ($\uppercase{X}=\text{Sn}$, \uppercase{T}i, \uppercase{G}e)},}\ }\href
  {\doibase 10.1103/PhysRevB.89.064401} {\bibfield  {journal} {\bibinfo
  {journal} {Phys. Rev. B}\ }\textbf {\bibinfo {volume} {89}},\ \bibinfo
  {pages} {064401} (\bibinfo {year} {2014})}\BibitemShut {NoStop}%
\bibitem [{\citenamefont {Dun}\ \emph {et~al.}(2015)\citenamefont {Dun},
  \citenamefont {Li}, \citenamefont {Freitas}, \citenamefont {Arrighi},
  \citenamefont {Dela~Cruz}, \citenamefont {Lee}, \citenamefont {Choi},
  \citenamefont {Cao}, \citenamefont {Silverstein}, \citenamefont {Wiebe},
  \citenamefont {Cheng},\ and\ \citenamefont {Zhou}}]{Dun15}%
  \BibitemOpen
  \bibfield  {author} {\bibinfo {author} {\bibfnamefont {Z.~L.}\ \bibnamefont
  {Dun}}, \bibinfo {author} {\bibfnamefont {X.}~\bibnamefont {Li}}, \bibinfo
  {author} {\bibfnamefont {R.~S.}\ \bibnamefont {Freitas}}, \bibinfo {author}
  {\bibfnamefont {E.}~\bibnamefont {Arrighi}}, \bibinfo {author} {\bibfnamefont
  {C.~R.}\ \bibnamefont {Dela~Cruz}}, \bibinfo {author} {\bibfnamefont
  {M.}~\bibnamefont {Lee}}, \bibinfo {author} {\bibfnamefont {E.~S.}\
  \bibnamefont {Choi}}, \bibinfo {author} {\bibfnamefont {H.~B.}\ \bibnamefont
  {Cao}}, \bibinfo {author} {\bibfnamefont {H.~J.}\ \bibnamefont
  {Silverstein}}, \bibinfo {author} {\bibfnamefont {C.~R.}\ \bibnamefont
  {Wiebe}}, \bibinfo {author} {\bibfnamefont {J.~G.}\ \bibnamefont {Cheng}}, \
  and\ \bibinfo {author} {\bibfnamefont {H.~D.}\ \bibnamefont {Zhou}},\
  }\bibfield  {title} {\enquote {\bibinfo {title} {Antiferromagnetic order in
  the pyrochlores ${R}_{2}\mathrm{Ge}_{2}\mathrm{O}_{7}\
  (\uppercase{R}=\mathrm{Er},\mathrm{Yb})$},}\ }\href {\doibase
  10.1103/PhysRevB.92.140407} {\bibfield  {journal} {\bibinfo  {journal} {Phys.
  Rev. B}\ }\textbf {\bibinfo {volume} {92}},\ \bibinfo {pages} {140407}
  (\bibinfo {year} {2015})}\BibitemShut {NoStop}%
\bibitem [{\citenamefont {Hallas}\ \emph
  {et~al.}(2016{\natexlab{b}})\citenamefont {Hallas}, \citenamefont {Gaudet},
  \citenamefont {Wilson}, \citenamefont {Munsie}, \citenamefont {Aczel},
  \citenamefont {Stone}, \citenamefont {Freitas}, \citenamefont
  {Arevalo-Lopez}, \citenamefont {Attfield}, \citenamefont {Tachibana},
  \citenamefont {Wiebe}, \citenamefont {Luke},\ and\ \citenamefont
  {Gaulin}}]{Hallas16a}%
  \BibitemOpen
  \bibfield  {author} {\bibinfo {author} {\bibfnamefont {A.~M.}\ \bibnamefont
  {Hallas}}, \bibinfo {author} {\bibfnamefont {J.}~\bibnamefont {Gaudet}},
  \bibinfo {author} {\bibfnamefont {M.~N.}\ \bibnamefont {Wilson}}, \bibinfo
  {author} {\bibfnamefont {T.~J.}\ \bibnamefont {Munsie}}, \bibinfo {author}
  {\bibfnamefont {A.~A.}\ \bibnamefont {Aczel}}, \bibinfo {author}
  {\bibfnamefont {M.~B.}\ \bibnamefont {Stone}}, \bibinfo {author}
  {\bibfnamefont {R.~S.}\ \bibnamefont {Freitas}}, \bibinfo {author}
  {\bibfnamefont {A.~M.}\ \bibnamefont {Arevalo-Lopez}}, \bibinfo {author}
  {\bibfnamefont {J.~P.}\ \bibnamefont {Attfield}}, \bibinfo {author}
  {\bibfnamefont {M.}~\bibnamefont {Tachibana}}, \bibinfo {author}
  {\bibfnamefont {C.~R.}\ \bibnamefont {Wiebe}}, \bibinfo {author}
  {\bibfnamefont {G.~M.}\ \bibnamefont {Luke}}, \ and\ \bibinfo {author}
  {\bibfnamefont {B.~D.}\ \bibnamefont {Gaulin}},\ }\bibfield  {title}
  {\enquote {\bibinfo {title} {\uppercase{XY} antiferromagnetic ground state in
  the effective $\uppercase{S}=\frac{1}{2}$ pyrochlore
  $\mathrm{Yb}_{2}\mathrm{Ge}_{2}\mathrm{O}_{7}$},}\ }\href {\doibase
  10.1103/PhysRevB.93.104405} {\bibfield  {journal} {\bibinfo  {journal} {Phys.
  Rev. B}\ }\textbf {\bibinfo {volume} {93}},\ \bibinfo {pages} {104405}
  (\bibinfo {year} {2016}{\natexlab{b}})}\BibitemShut {NoStop}%
\bibitem [{\citenamefont {Hallas}\ \emph
  {et~al.}(2016{\natexlab{c}})\citenamefont {Hallas}, \citenamefont {Gaudet},
  \citenamefont {Butch}, \citenamefont {Tachibana}, \citenamefont {Freitas},
  \citenamefont {Luke}, \citenamefont {Wiebe},\ and\ \citenamefont
  {Gaulin}}]{Hallas16}%
  \BibitemOpen
  \bibfield  {author} {\bibinfo {author} {\bibfnamefont {A.~M.}\ \bibnamefont
  {Hallas}}, \bibinfo {author} {\bibfnamefont {J.}~\bibnamefont {Gaudet}},
  \bibinfo {author} {\bibfnamefont {N.~P.}\ \bibnamefont {Butch}}, \bibinfo
  {author} {\bibfnamefont {M.}~\bibnamefont {Tachibana}}, \bibinfo {author}
  {\bibfnamefont {R.~S.}\ \bibnamefont {Freitas}}, \bibinfo {author}
  {\bibfnamefont {G.~M.}\ \bibnamefont {Luke}}, \bibinfo {author}
  {\bibfnamefont {C.~R.}\ \bibnamefont {Wiebe}}, \ and\ \bibinfo {author}
  {\bibfnamefont {B.~D.}\ \bibnamefont {Gaulin}},\ }\bibfield  {title}
  {\enquote {\bibinfo {title} {Universal dynamic magnetism in \uppercase{Y}b
  pyrochlores with disparate ground states},}\ }\href {\doibase
  10.1103/PhysRevB.93.100403} {\bibfield  {journal} {\bibinfo  {journal} {Phys.
  Rev. B}\ }\textbf {\bibinfo {volume} {93}},\ \bibinfo {pages} {100403}
  (\bibinfo {year} {2016}{\natexlab{c}})}\BibitemShut {NoStop}%
\bibitem [{\citenamefont {Bertin}\ \emph {et~al.}(2012)\citenamefont {Bertin},
  \citenamefont {Chapuis}, \citenamefont {{Dalmas de R\'eotier}},\ and\
  \citenamefont {Yaouanc}}]{Bertin12}%
  \BibitemOpen
  \bibfield  {author} {\bibinfo {author} {\bibfnamefont {A.}~\bibnamefont
  {Bertin}}, \bibinfo {author} {\bibfnamefont {Y.}~\bibnamefont {Chapuis}},
  \bibinfo {author} {\bibfnamefont {P.}~\bibnamefont {{Dalmas de R\'eotier}}},
  \ and\ \bibinfo {author} {\bibfnamefont {A.}~\bibnamefont {Yaouanc}},\
  }\bibfield  {title} {\enquote {\bibinfo {title} {Crystal electric field at
  the rare-earth ion in the
  \uppercase{R}$_2$\uppercase{T}i$_2$\uppercase{O}$_7$ pyrochlore compounds},}\
  }\href {http://stacks.iop.org/0953-8984/24/i=25/a=256003} {\bibfield
  {journal} {\bibinfo  {journal} {J. Phys.: Condens. Matter}\ }\textbf
  {\bibinfo {volume} {24}},\ \bibinfo {pages} {256003} (\bibinfo {year}
  {2012})}\BibitemShut {NoStop}%
\bibitem [{\citenamefont {Lau}\ \emph {et~al.}(2005)\citenamefont {Lau},
  \citenamefont {Freitas}, \citenamefont {Ueland}, \citenamefont {Schiffer},\
  and\ \citenamefont {Cava}}]{Lau05}%
  \BibitemOpen
  \bibfield  {author} {\bibinfo {author} {\bibfnamefont {G.~C.}\ \bibnamefont
  {Lau}}, \bibinfo {author} {\bibfnamefont {R.~S.}\ \bibnamefont {Freitas}},
  \bibinfo {author} {\bibfnamefont {B.~G.}\ \bibnamefont {Ueland}}, \bibinfo
  {author} {\bibfnamefont {P.}~\bibnamefont {Schiffer}}, \ and\ \bibinfo
  {author} {\bibfnamefont {R.~J.}\ \bibnamefont {Cava}},\ }\bibfield  {title}
  {\enquote {\bibinfo {title} {Geometrical magnetic frustration in rare-earth
  chalcogenide spinels},}\ }\href {\doibase 10.1103/PhysRevB.72.054411}
  {\bibfield  {journal} {\bibinfo  {journal} {Phys. Rev. B}\ }\textbf {\bibinfo
  {volume} {72}},\ \bibinfo {pages} {054411} (\bibinfo {year}
  {2005})}\BibitemShut {NoStop}%
\bibitem [{\citenamefont {Yaouanc}\ \emph {et~al.}(2011)\citenamefont
  {Yaouanc}, \citenamefont {{Dalmas de R\'eotier}}, \citenamefont {Marin},\
  and\ \citenamefont {Glazkov}}]{Yaouanc11c}%
  \BibitemOpen
  \bibfield  {author} {\bibinfo {author} {\bibfnamefont {A.}~\bibnamefont
  {Yaouanc}}, \bibinfo {author} {\bibfnamefont {P.}~\bibnamefont {{Dalmas de
  R\'eotier}}}, \bibinfo {author} {\bibfnamefont {C.}~\bibnamefont {Marin}}, \
  and\ \bibinfo {author} {\bibfnamefont {V.}~\bibnamefont {Glazkov}},\
  }\bibfield  {title} {\enquote {\bibinfo {title} {Single-crystal versus
  polycrystalline samples of magnetically frustrated
  \uppercase{Y}b$_2$\uppercase{T}i$_2$\uppercase{O}$_7$: Specific heat
  results},}\ }\href {\doibase 10.1103/PhysRevB.84.172408} {\bibfield
  {journal} {\bibinfo  {journal} {Phys. Rev. B}\ }\textbf {\bibinfo {volume}
  {84}},\ \bibinfo {pages} {172408} (\bibinfo {year} {2011})}\BibitemShut
  {NoStop}%
\bibitem [{\citenamefont {Lago}\ \emph {et~al.}(2010)\citenamefont {Lago},
  \citenamefont {\ifmmode \check{Z}\else
  \v{Z}\fi{}ivkovi\ifmmode~\acute{c}\else \'{c}\fi{}}, \citenamefont {Malkin},
  \citenamefont {Rodriguez~Fernandez}, \citenamefont {Ghigna}, \citenamefont
  {Dalmas~de R\'eotier}, \citenamefont {Yaouanc},\ and\ \citenamefont
  {Rojo}}]{Lago10}%
  \BibitemOpen
  \bibfield  {author} {\bibinfo {author} {\bibfnamefont {J.}~\bibnamefont
  {Lago}}, \bibinfo {author} {\bibfnamefont {I.}~\bibnamefont {\ifmmode
  \check{Z}\else \v{Z}\fi{}ivkovi\ifmmode~\acute{c}\else \'{c}\fi{}}}, \bibinfo
  {author} {\bibfnamefont {B.~Z.}\ \bibnamefont {Malkin}}, \bibinfo {author}
  {\bibfnamefont {J.}~\bibnamefont {Rodriguez~Fernandez}}, \bibinfo {author}
  {\bibfnamefont {P.}~\bibnamefont {Ghigna}}, \bibinfo {author} {\bibfnamefont
  {P.}~\bibnamefont {Dalmas~de R\'eotier}}, \bibinfo {author} {\bibfnamefont
  {A.}~\bibnamefont {Yaouanc}}, \ and\ \bibinfo {author} {\bibfnamefont
  {T.}~\bibnamefont {Rojo}},\ }\bibfield  {title} {\enquote {\bibinfo {title}
  {$\mathrm{CdEr}_{2}\mathrm{Se}_{4}$: A new erbium spin ice system in a spinel
  structure},}\ }\href {\doibase 10.1103/PhysRevLett.104.247203} {\bibfield
  {journal} {\bibinfo  {journal} {Phys. Rev. Lett.}\ }\textbf {\bibinfo
  {volume} {104}},\ \bibinfo {pages} {247203} (\bibinfo {year}
  {2010})}\BibitemShut {NoStop}%
\bibitem [{\citenamefont {Legros}\ \emph {et~al.}(2015)\citenamefont {Legros},
  \citenamefont {Ryan}, \citenamefont {Dalmas~de R{\'e}otier}, \citenamefont
  {Yaouanc},\ and\ \citenamefont {Marin}}]{Legros15}%
  \BibitemOpen
  \bibfield  {author} {\bibinfo {author} {\bibfnamefont {Ana{\"e}lle}\
  \bibnamefont {Legros}}, \bibinfo {author} {\bibfnamefont {D.~H.}\
  \bibnamefont {Ryan}}, \bibinfo {author} {\bibfnamefont {P.}~\bibnamefont
  {Dalmas~de R{\'e}otier}}, \bibinfo {author} {\bibfnamefont {A.}~\bibnamefont
  {Yaouanc}}, \ and\ \bibinfo {author} {\bibfnamefont {C.}~\bibnamefont
  {Marin}},\ }\bibfield  {title} {\enquote {\bibinfo {title}
  {$^{166}$\uppercase{E}r \uppercase{M}{\"o}ssbauer spectroscopy study of
  magnetic ordering in a spinel-based potential spin-ice system:
  \uppercase{C}d\uppercase{E}r$_2$\uppercase{S}$_4$},}\ }\href {\doibase
  http://dx.doi.org/10.1063/1.4906182} {\bibfield  {journal} {\bibinfo
  {journal} {J. Appl. Phys.}\ }\textbf {\bibinfo {volume} {117}},\ \bibinfo
  {eid} {17C701} (\bibinfo {year} {2015})}\BibitemShut {NoStop}%
\bibitem [{\citenamefont {Yaouanc}\ \emph {et~al.}(2015)\citenamefont
  {Yaouanc}, \citenamefont {Dalmas~de R\'eotier}, \citenamefont {Bertin},
  \citenamefont {Marin}, \citenamefont {Lhotel}, \citenamefont {Amato},\ and\
  \citenamefont {Baines}}]{Yaouanc15}%
  \BibitemOpen
  \bibfield  {author} {\bibinfo {author} {\bibfnamefont {A.}~\bibnamefont
  {Yaouanc}}, \bibinfo {author} {\bibfnamefont {P.}~\bibnamefont {Dalmas~de
  R\'eotier}}, \bibinfo {author} {\bibfnamefont {A.}~\bibnamefont {Bertin}},
  \bibinfo {author} {\bibfnamefont {C.}~\bibnamefont {Marin}}, \bibinfo
  {author} {\bibfnamefont {E.}~\bibnamefont {Lhotel}}, \bibinfo {author}
  {\bibfnamefont {A.}~\bibnamefont {Amato}}, \ and\ \bibinfo {author}
  {\bibfnamefont {C.}~\bibnamefont {Baines}},\ }\bibfield  {title} {\enquote
  {\bibinfo {title} {Evidence for unidimensional low-energy excitations as the
  origin of persistent spin dynamics in geometrically frustrated magnets},}\
  }\href {\doibase 10.1103/PhysRevB.91.104427} {\bibfield  {journal} {\bibinfo
  {journal} {Phys. Rev. B}\ }\textbf {\bibinfo {volume} {91}},\ \bibinfo
  {pages} {104427} (\bibinfo {year} {2015})}\BibitemShut {NoStop}%
\bibitem [{\citenamefont {Higo}\ \emph {et~al.}(2017)\citenamefont {Higo},
  \citenamefont {Iritani}, \citenamefont {Halim}, \citenamefont {Higemoto},
  \citenamefont {Ito}, \citenamefont {Kuga}, \citenamefont {Kimura},\ and\
  \citenamefont {Nakatsuji}}]{Higo17}%
  \BibitemOpen
  \bibfield  {author} {\bibinfo {author} {\bibfnamefont {Tomoya}\ \bibnamefont
  {Higo}}, \bibinfo {author} {\bibfnamefont {Kensuke}\ \bibnamefont {Iritani}},
  \bibinfo {author} {\bibfnamefont {Mario}\ \bibnamefont {Halim}}, \bibinfo
  {author} {\bibfnamefont {Wataru}\ \bibnamefont {Higemoto}}, \bibinfo {author}
  {\bibfnamefont {Takashi~U.}\ \bibnamefont {Ito}}, \bibinfo {author}
  {\bibfnamefont {Kentaro}\ \bibnamefont {Kuga}}, \bibinfo {author}
  {\bibfnamefont {Kenta}\ \bibnamefont {Kimura}}, \ and\ \bibinfo {author}
  {\bibfnamefont {Satoru}\ \bibnamefont {Nakatsuji}},\ }\bibfield  {title}
  {\enquote {\bibinfo {title} {Frustrated magnetism in the
  \uppercase{H}eisenberg pyrochlore antiferromagnets
  ${A}\mathrm{Yb}_{2}{X}_{4}$ (${A}$ = \uppercase{C}d, \uppercase{M}g;
  $\uppercase{X}=\mathrm{S}$, $\mathrm{Se}$)},}\ }\href {\doibase
  10.1103/PhysRevB.95.174443} {\bibfield  {journal} {\bibinfo  {journal} {Phys.
  Rev. B}\ }\textbf {\bibinfo {volume} {95}},\ \bibinfo {pages} {174443}
  (\bibinfo {year} {2017})}\BibitemShut {NoStop}%
\bibitem [{\citenamefont {Yaouanc}\ \emph {et~al.}(2013)\citenamefont
  {Yaouanc}, \citenamefont {Dalmas~de R\'eotier}, \citenamefont {Bonville},
  \citenamefont {Hodges}, \citenamefont {Glazkov}, \citenamefont {Keller},
  \citenamefont {Sikolenko}, \citenamefont {Bartkowiak}, \citenamefont {Amato},
  \citenamefont {Baines}, \citenamefont {King}, \citenamefont {Gubbens},\ and\
  \citenamefont {Forget}}]{Yaouanc13}%
  \BibitemOpen
  \bibfield  {author} {\bibinfo {author} {\bibfnamefont {A.}~\bibnamefont
  {Yaouanc}}, \bibinfo {author} {\bibfnamefont {P.}~\bibnamefont {Dalmas~de
  R\'eotier}}, \bibinfo {author} {\bibfnamefont {P.}~\bibnamefont {Bonville}},
  \bibinfo {author} {\bibfnamefont {J.~A.}\ \bibnamefont {Hodges}}, \bibinfo
  {author} {\bibfnamefont {V.}~\bibnamefont {Glazkov}}, \bibinfo {author}
  {\bibfnamefont {L.}~\bibnamefont {Keller}}, \bibinfo {author} {\bibfnamefont
  {V.}~\bibnamefont {Sikolenko}}, \bibinfo {author} {\bibfnamefont
  {M.}~\bibnamefont {Bartkowiak}}, \bibinfo {author} {\bibfnamefont
  {A.}~\bibnamefont {Amato}}, \bibinfo {author} {\bibfnamefont
  {C.}~\bibnamefont {Baines}}, \bibinfo {author} {\bibfnamefont {P.~J.~C.}\
  \bibnamefont {King}}, \bibinfo {author} {\bibfnamefont {P.~C.~M.}\
  \bibnamefont {Gubbens}}, \ and\ \bibinfo {author} {\bibfnamefont
  {A.}~\bibnamefont {Forget}},\ }\bibfield  {title} {\enquote {\bibinfo {title}
  {Dynamical splayed ferromagnetic ground state in the quantum spin ice
  \uppercase{Y}b$_2$\uppercase{S}n$_2$\uppercase{O}$_7$},}\ }\href {\doibase
  10.1103/PhysRevLett.110.127207} {\bibfield  {journal} {\bibinfo  {journal}
  {Phys. Rev. Lett.}\ }\textbf {\bibinfo {volume} {110}},\ \bibinfo {pages}
  {127207} (\bibinfo {year} {2013})}\BibitemShut {NoStop}%
\bibitem [{\citenamefont {Yaouanc}\ \emph {et~al.}(2016)\citenamefont
  {Yaouanc}, \citenamefont {{Dalmas de R\'eotier}}, \citenamefont {Keller},
  \citenamefont {Roessli},\ and\ \citenamefont {Forget}}]{Yaouanc16}%
  \BibitemOpen
  \bibfield  {author} {\bibinfo {author} {\bibfnamefont {A.}~\bibnamefont
  {Yaouanc}}, \bibinfo {author} {\bibfnamefont {P.}~\bibnamefont {{Dalmas de
  R\'eotier}}}, \bibinfo {author} {\bibfnamefont {L.}~\bibnamefont {Keller}},
  \bibinfo {author} {\bibfnamefont {B.}~\bibnamefont {Roessli}}, \ and\
  \bibinfo {author} {\bibfnamefont {A.}~\bibnamefont {Forget}},\ }\bibfield
  {title} {\enquote {\bibinfo {title} {A novel type of splayed ferromagnetic
  order observed in \uppercase{Y}b$_2$\uppercase{T}i$_2$\uppercase{O}$_7$},}\
  }\href {http://stacks.iop.org/0953-8984/28/i=42/a=426002} {\bibfield
  {journal} {\bibinfo  {journal} {J.\ Phys.:\ Condens.\ Matter}\ }\textbf
  {\bibinfo {volume} {28}},\ \bibinfo {pages} {426002} (\bibinfo {year}
  {2016})}\BibitemShut {NoStop}%
\bibitem [{\citenamefont {{Dalmas de R\'eotier}}\ \emph
  {et~al.}(2016)\citenamefont {{Dalmas de R\'eotier}}, \citenamefont
  {Maisuradze},\ and\ \citenamefont {Yaouanc}}]{Dalmas16a}%
  \BibitemOpen
  \bibfield  {author} {\bibinfo {author} {\bibfnamefont {P.}~\bibnamefont
  {{Dalmas de R\'eotier}}}, \bibinfo {author} {\bibfnamefont {A.}~\bibnamefont
  {Maisuradze}}, \ and\ \bibinfo {author} {\bibfnamefont {A.}~\bibnamefont
  {Yaouanc}},\ }\bibfield  {title} {\enquote {\bibinfo {title} {Recent
  $\mu$\uppercase{SR} studies of insulating rare-earth pyrochlore magnets},}\
  }\href {\doibase 10.7566/JPSJ.85.091010} {\bibfield  {journal} {\bibinfo
  {journal} {J. Phys. Soc. Jpn.}\ }\textbf {\bibinfo {volume} {85}},\ \bibinfo
  {pages} {091010} (\bibinfo {year} {2016})}\BibitemShut {NoStop}%
\bibitem [{\citenamefont {Dalmas~de R\'eotier}\ \emph
  {et~al.}(2017)\citenamefont {Dalmas~de R\'eotier}, \citenamefont {Yaouanc},
  \citenamefont {Maisuradze}, \citenamefont {Bertin}, \citenamefont {Baker},
  \citenamefont {Hillier},\ and\ \citenamefont {Forget}}]{Dalmas17a}%
  \BibitemOpen
  \bibfield  {author} {\bibinfo {author} {\bibfnamefont {P.}~\bibnamefont
  {Dalmas~de R\'eotier}}, \bibinfo {author} {\bibfnamefont {A.}~\bibnamefont
  {Yaouanc}}, \bibinfo {author} {\bibfnamefont {A.}~\bibnamefont {Maisuradze}},
  \bibinfo {author} {\bibfnamefont {A.}~\bibnamefont {Bertin}}, \bibinfo
  {author} {\bibfnamefont {P.~J.}\ \bibnamefont {Baker}}, \bibinfo {author}
  {\bibfnamefont {A.~D.}\ \bibnamefont {Hillier}}, \ and\ \bibinfo {author}
  {\bibfnamefont {A.}~\bibnamefont {Forget}},\ }\bibfield  {title} {\enquote
  {\bibinfo {title} {Slow spin tunneling in the paramagnetic phase of the
  pyrochlore \uppercase{N}d$_2$\uppercase{S}n$_2$\uppercase{O}$_7$},}\ }\href
  {\doibase 10.1103/PhysRevB.95.134420} {\bibfield  {journal} {\bibinfo
  {journal} {Phys. Rev. B}\ }\textbf {\bibinfo {volume} {95}},\ \bibinfo
  {pages} {134420} (\bibinfo {year} {2017})}\BibitemShut {NoStop}%
\bibitem [{\citenamefont {{Dalmas de R\'eotier}}\ \emph
  {et~al.}(2003)\citenamefont {{Dalmas de R\'eotier}}, \citenamefont {Yaouanc},
  \citenamefont {Gubbens}, \citenamefont {Kaiser}, \citenamefont {Baines},\
  and\ \citenamefont {King}}]{Dalmas03}%
  \BibitemOpen
  \bibfield  {author} {\bibinfo {author} {\bibfnamefont {P.}~\bibnamefont
  {{Dalmas de R\'eotier}}}, \bibinfo {author} {\bibfnamefont {A.}~\bibnamefont
  {Yaouanc}}, \bibinfo {author} {\bibfnamefont {P.~C.~M.}\ \bibnamefont
  {Gubbens}}, \bibinfo {author} {\bibfnamefont {C.~T.}\ \bibnamefont {Kaiser}},
  \bibinfo {author} {\bibfnamefont {C.}~\bibnamefont {Baines}}, \ and\ \bibinfo
  {author} {\bibfnamefont {P.~J.~C.}\ \bibnamefont {King}},\ }\bibfield
  {title} {\enquote {\bibinfo {title} {Absence of magnetic order in
  $\mathrm{Y}\mathrm{b}_{3}\mathrm{G}\mathrm{a}_{5}\mathrm{O}_{12}$: Relation
  between phase transition and entropy in geometrically frustrated
  materials},}\ }\href {\doibase 10.1103/PhysRevLett.91.167201} {\bibfield
  {journal} {\bibinfo  {journal} {Phys. Rev. Lett.}\ }\textbf {\bibinfo
  {volume} {91}},\ \bibinfo {pages} {167201} (\bibinfo {year}
  {2003})}\BibitemShut {NoStop}%
\bibitem [{\citenamefont {Dalmas~de R\'eotier}\ \emph
  {et~al.}(2012)\citenamefont {Dalmas~de R\'eotier}, \citenamefont {Yaouanc},
  \citenamefont {Chapuis}, \citenamefont {Curnoe}, \citenamefont {Grenier},
  \citenamefont {Ressouche}, \citenamefont {Marin}, \citenamefont {Lago},
  \citenamefont {Baines},\ and\ \citenamefont {Giblin}}]{Dalmas12a}%
  \BibitemOpen
  \bibfield  {author} {\bibinfo {author} {\bibfnamefont {P.}~\bibnamefont
  {Dalmas~de R\'eotier}}, \bibinfo {author} {\bibfnamefont {A.}~\bibnamefont
  {Yaouanc}}, \bibinfo {author} {\bibfnamefont {Y.}~\bibnamefont {Chapuis}},
  \bibinfo {author} {\bibfnamefont {S.~H.}\ \bibnamefont {Curnoe}}, \bibinfo
  {author} {\bibfnamefont {B.}~\bibnamefont {Grenier}}, \bibinfo {author}
  {\bibfnamefont {E.}~\bibnamefont {Ressouche}}, \bibinfo {author}
  {\bibfnamefont {C.}~\bibnamefont {Marin}}, \bibinfo {author} {\bibfnamefont
  {J.}~\bibnamefont {Lago}}, \bibinfo {author} {\bibfnamefont {C.}~\bibnamefont
  {Baines}}, \ and\ \bibinfo {author} {\bibfnamefont {S.~R.}\ \bibnamefont
  {Giblin}},\ }\bibfield  {title} {\enquote {\bibinfo {title} {Magnetic order,
  magnetic correlations and spin dynamics in the pyrochlore antiferromagnet
  \uppercase{E}r$_2$\uppercase{T}i$_2$\uppercase{O}$_7$},}\ }\href {\doibase
  10.1103/PhysRevB.86.104424} {\bibfield  {journal} {\bibinfo  {journal} {Phys.
  Rev. B}\ }\textbf {\bibinfo {volume} {86}},\ \bibinfo {pages} {104424}
  (\bibinfo {year} {2012})}\BibitemShut {NoStop}%
\bibitem [{\citenamefont {Ramirez}(2001)}]{Ramirez01}%
  \BibitemOpen
  \bibfield  {author} {\bibinfo {author} {\bibfnamefont {A.~P.}\ \bibnamefont
  {Ramirez}},\ }in\ \href@noop {} {\emph {\bibinfo {booktitle} {Handbook of
  Magnetic Materials}}},\ Vol.~\bibinfo {volume} {13},\ \bibinfo {editor}
  {edited by\ \bibinfo {editor} {\bibfnamefont {K.~H.~J.}\ \bibnamefont
  {Buschow}}}\ (\bibinfo  {publisher} {Elsevier},\ \bibinfo {address}
  {Amsterdam},\ \bibinfo {year} {2001})\BibitemShut {NoStop}%
\bibitem [{\citenamefont {Ben-Dor}\ and\ \citenamefont
  {Shilo}(1980)}]{Ben-Dor80}%
  \BibitemOpen
  \bibfield  {author} {\bibinfo {author} {\bibfnamefont {L.}~\bibnamefont
  {Ben-Dor}}\ and\ \bibinfo {author} {\bibfnamefont {I.}~\bibnamefont
  {Shilo}},\ }\bibfield  {title} {\enquote {\bibinfo {title} {Structure and
  magnetic properties of sulfides of the type
  \uppercase{C}d\textit{RE}$_2$\uppercase{S}$_4$ and
  \uppercase{M}g(\uppercase{G}d$_x$\uppercase{Y}b$_{1-x}$)$_2$\uppercase{S}$_4$},}\
  }\href {https://doi.org/10.1016/0022-4596(80)90504-6} {\bibfield  {journal}
  {\bibinfo  {journal} {J. Solid State Chem.}\ }\textbf {\bibinfo {volume}
  {35}},\ \bibinfo {pages} {278} (\bibinfo {year} {1980})}\BibitemShut
  {NoStop}%
\bibitem [{\citenamefont {Suchow}\ and\ \citenamefont
  {Stemple}(1964)}]{Suchow64}%
  \BibitemOpen
  \bibfield  {author} {\bibinfo {author} {\bibfnamefont {L.}~\bibnamefont
  {Suchow}}\ and\ \bibinfo {author} {\bibfnamefont {N.~R.}\ \bibnamefont
  {Stemple}},\ }\bibfield  {title} {\enquote {\bibinfo {title} {Fluorescent
  rare earths in semiconducting thiospinels},}\ }\href {\doibase
  10.1149/1.2426081} {\bibfield  {journal} {\bibinfo  {journal} {J.
  Electrochem. Soc.}\ }\textbf {\bibinfo {volume} {111}},\ \bibinfo {pages}
  {191} (\bibinfo {year} {1964})}\BibitemShut {NoStop}%
\bibitem [{\citenamefont {Rodriguez-Carvajal}(1993)}]{Rodriguez93}%
  \BibitemOpen
  \bibfield  {author} {\bibinfo {author} {\bibfnamefont {J.}~\bibnamefont
  {Rodriguez-Carvajal}},\ }\bibfield  {title} {\enquote {\bibinfo {title}
  {Recent advances in magnetic structure determination by neutron powder
  diffraction},}\ }\href {\doibase 10.1016/0921-4526(93)90108-I} {\bibfield
  {journal} {\bibinfo  {journal} {Physica B}\ }\textbf {\bibinfo {volume}
  {192}},\ \bibinfo {pages} {55} (\bibinfo {year} {1993})}\BibitemShut
  {NoStop}%
\bibitem [{\citenamefont {Wills}\ \emph {et~al.}(2006)\citenamefont {Wills},
  \citenamefont {Zhitomirsky}, \citenamefont {Canals}, \citenamefont {Sanchez},
  \citenamefont {Bonville}, \citenamefont {{Dalmas de R\'eotier}},\ and\
  \citenamefont {Yaouanc}}]{Wills06}%
  \BibitemOpen
  \bibfield  {author} {\bibinfo {author} {\bibfnamefont {A~S}\ \bibnamefont
  {Wills}}, \bibinfo {author} {\bibfnamefont {M~E}\ \bibnamefont
  {Zhitomirsky}}, \bibinfo {author} {\bibfnamefont {B}~\bibnamefont {Canals}},
  \bibinfo {author} {\bibfnamefont {J~P}\ \bibnamefont {Sanchez}}, \bibinfo
  {author} {\bibfnamefont {P}~\bibnamefont {Bonville}}, \bibinfo {author}
  {\bibfnamefont {P}~\bibnamefont {{Dalmas de R\'eotier}}}, \ and\ \bibinfo
  {author} {\bibfnamefont {A}~\bibnamefont {Yaouanc}},\ }\bibfield  {title}
  {\enquote {\bibinfo {title} {Magnetic ordering in
  \uppercase{G}d$_2$\uppercase{S}n$_2$\uppercase{O}$_7$: the archetypal
  \uppercase{H}eisenberg pyrochlore antiferromagnet},}\ }\href
  {http://stacks.iop.org/0953-8984/18/i=3/a=L02} {\bibfield  {journal}
  {\bibinfo  {journal} {J. Phys.: Condens. Matter}\ }\textbf {\bibinfo {volume}
  {18}},\ \bibinfo {pages} {L37--L42} (\bibinfo {year} {2006})}\BibitemShut
  {NoStop}%
\bibitem [{\citenamefont {Poole}\ \emph {et~al.}(2007)\citenamefont {Poole},
  \citenamefont {Wills},\ and\ \citenamefont {Leli\`evre-Berna}}]{Poole07}%
  \BibitemOpen
  \bibfield  {author} {\bibinfo {author} {\bibfnamefont {A.}~\bibnamefont
  {Poole}}, \bibinfo {author} {\bibfnamefont {A.~S.}\ \bibnamefont {Wills}}, \
  and\ \bibinfo {author} {\bibfnamefont {E.}~\bibnamefont {Leli\`evre-Berna}},\
  }\bibfield  {title} {\enquote {\bibinfo {title} {Magnetic ordering in the
  \uppercase{XY} pyrochlore antiferromagnet
  \uppercase{E}r$_2$\uppercase{T}i$_2$\uppercase{O}$_7$: a spherical neutron
  polarimetry study},}\ }\href
  {http://stacks.iop.org/0953-8984/19/i=45/a=452201} {\bibfield  {journal}
  {\bibinfo  {journal} {J. Phys.: Condens. Matter}\ }\textbf {\bibinfo {volume}
  {19}},\ \bibinfo {pages} {452201} (\bibinfo {year} {2007})}\BibitemShut
  {NoStop}%
\bibitem [{\citenamefont {Champion}\ \emph {et~al.}(2003)\citenamefont
  {Champion}, \citenamefont {Harris}, \citenamefont {Holdsworth}, \citenamefont
  {Wills}, \citenamefont {Balakrishnan}, \citenamefont {Bramwell},
  \citenamefont {\ifmmode \check{C}\else \v{C}\fi{}i\ifmmode~\check{z}\else
  \v{z}\fi{}m\'ar}, \citenamefont {Fennell}, \citenamefont {Gardner},
  \citenamefont {Lago}, \citenamefont {McMorrow}, \citenamefont
  {Orend\'a\ifmmode~\check{c}\else \v{c}\fi{}}, \citenamefont
  {Orend\'a\ifmmode~\check{c}\else \v{c}\fi{}ov\'a}, \citenamefont {{McK.
  Paul}}, \citenamefont {Smith}, \citenamefont {Telling},\ and\ \citenamefont
  {Wildes}}]{Champion03}%
  \BibitemOpen
  \bibfield  {author} {\bibinfo {author} {\bibfnamefont {J.~D.~M.}\
  \bibnamefont {Champion}}, \bibinfo {author} {\bibfnamefont {M.~J.}\
  \bibnamefont {Harris}}, \bibinfo {author} {\bibfnamefont {P.~C.~W.}\
  \bibnamefont {Holdsworth}}, \bibinfo {author} {\bibfnamefont {A.~S.}\
  \bibnamefont {Wills}}, \bibinfo {author} {\bibfnamefont {G.}~\bibnamefont
  {Balakrishnan}}, \bibinfo {author} {\bibfnamefont {S.~T.}\ \bibnamefont
  {Bramwell}}, \bibinfo {author} {\bibfnamefont {E.}~\bibnamefont {\ifmmode
  \check{C}\else \v{C}\fi{}i\ifmmode~\check{z}\else \v{z}\fi{}m\'ar}}, \bibinfo
  {author} {\bibfnamefont {T.}~\bibnamefont {Fennell}}, \bibinfo {author}
  {\bibfnamefont {J.~S.}\ \bibnamefont {Gardner}}, \bibinfo {author}
  {\bibfnamefont {J.}~\bibnamefont {Lago}}, \bibinfo {author} {\bibfnamefont
  {D.~F.}\ \bibnamefont {McMorrow}}, \bibinfo {author} {\bibfnamefont
  {M.}~\bibnamefont {Orend\'a\ifmmode~\check{c}\else \v{c}\fi{}}}, \bibinfo
  {author} {\bibfnamefont {A.}~\bibnamefont {Orend\'a\ifmmode~\check{c}\else
  \v{c}\fi{}ov\'a}}, \bibinfo {author} {\bibfnamefont {D.~McK.}\ \bibnamefont
  {{McK. Paul}}}, \bibinfo {author} {\bibfnamefont {R.~I.}\ \bibnamefont
  {Smith}}, \bibinfo {author} {\bibfnamefont {M.~T.~F.}\ \bibnamefont
  {Telling}}, \ and\ \bibinfo {author} {\bibfnamefont {A.}~\bibnamefont
  {Wildes}},\ }\bibfield  {title} {\enquote {\bibinfo {title}
  {\uppercase{E}r$_2$\uppercase{T}i$_2$\uppercase{O}$_7$: Evidence of quantum
  order by disorder in a frustrated antiferromagnet},}\ }\href {\doibase
  10.1103/PhysRevB.68.020401} {\bibfield  {journal} {\bibinfo  {journal} {Phys.
  Rev. B}\ }\textbf {\bibinfo {volume} {68}},\ \bibinfo {pages} {020401}
  (\bibinfo {year} {2003})}\BibitemShut {NoStop}%
\bibitem [{\citenamefont {Yaouanc}\ and\ \citenamefont {{Dalmas de
  R\'eotier}}(2011)}]{Yaouanc11}%
  \BibitemOpen
  \bibfield  {author} {\bibinfo {author} {\bibfnamefont {A.}~\bibnamefont
  {Yaouanc}}\ and\ \bibinfo {author} {\bibfnamefont {P.}~\bibnamefont {{Dalmas
  de R\'eotier}}},\ }\href@noop {} {\emph {\bibinfo {title} {Muon Spin
  Rotation, Relaxation, and Resonance: Applications to Condensed Matter}}}\
  (\bibinfo  {publisher} {Oxford University Press},\ \bibinfo {address}
  {Oxford},\ \bibinfo {year} {2011})\BibitemShut {NoStop}%
\bibitem [{\citenamefont {Maisuradze}\ \emph {et~al.}(2015)\citenamefont
  {Maisuradze}, \citenamefont {Dalmas~de R\'eotier}, \citenamefont {Yaouanc},
  \citenamefont {Forget}, \citenamefont {Baines},\ and\ \citenamefont
  {King}}]{Maisuradze15}%
  \BibitemOpen
  \bibfield  {author} {\bibinfo {author} {\bibfnamefont {A.}~\bibnamefont
  {Maisuradze}}, \bibinfo {author} {\bibfnamefont {P.}~\bibnamefont {Dalmas~de
  R\'eotier}}, \bibinfo {author} {\bibfnamefont {A.}~\bibnamefont {Yaouanc}},
  \bibinfo {author} {\bibfnamefont {A.}~\bibnamefont {Forget}}, \bibinfo
  {author} {\bibfnamefont {C.}~\bibnamefont {Baines}}, \ and\ \bibinfo {author}
  {\bibfnamefont {P.~J.~C.}\ \bibnamefont {King}},\ }\bibfield  {title}
  {\enquote {\bibinfo {title} {Anomalously slow spin dynamics and short-range
  correlations in the quantum spin ice systems
  $\mathrm{Yb}_{2}\mathrm{Ti}_{2}\mathrm{O}_{7}$ and
  $\mathrm{Yb}_{2}\mathrm{Sn}_{2}\mathrm{O}_{7}$},}\ }\href {\doibase
  10.1103/PhysRevB.92.094424} {\bibfield  {journal} {\bibinfo  {journal} {Phys.
  Rev. B}\ }\textbf {\bibinfo {volume} {92}},\ \bibinfo {pages} {094424}
  (\bibinfo {year} {2015})}\BibitemShut {NoStop}%
\bibitem [{Note1()}]{Note1}%
  \BibitemOpen
  \bibinfo {note} {Electron spin resonance measurements suggest the
  CdYb$_2$S$_4$ magnetic structure to be described by $\psi _2$ rather than
  $\psi _3$ \cite {Yoshizawa15}. Further neutron diffraction, e.g. in an
  applied magnetic field, would be welcome for a confirmation.}\BibitemShut
  {Stop}%
\bibitem [{not()}]{note_Yb2Ti2O7_Yb2Sn2O7}%
  \BibitemOpen
  \href@noop {} {}\bibinfo {note} {Yb$_2$Ti$_2$O$_7$ and Yb$_2$Sn$_2$O$_7$ have
  been shown to exhibit two different variants of splayed ferromagnetic order,
  distinguished by different magnetic components perpendicular to their easy
  axes \cite{Yaouanc13,Yaouanc16} although this point is disputed
  \cite{Gaudet16}.}\BibitemShut {Stop}%
\bibitem [{\citenamefont {Perez-Mato}\ \emph {et~al.}(2015)\citenamefont
  {Perez-Mato}, \citenamefont {Gallego}, \citenamefont {Tasci}, \citenamefont
  {Elcoro}, \citenamefont {{de la Flor}},\ and\ \citenamefont
  {Aroyo}}]{Perez15}%
  \BibitemOpen
  \bibfield  {author} {\bibinfo {author} {\bibfnamefont {J.~M.}\ \bibnamefont
  {Perez-Mato}}, \bibinfo {author} {\bibfnamefont {S.~V.}\ \bibnamefont
  {Gallego}}, \bibinfo {author} {\bibfnamefont {E.~S.}\ \bibnamefont {Tasci}},
  \bibinfo {author} {\bibfnamefont {L.}~\bibnamefont {Elcoro}}, \bibinfo
  {author} {\bibfnamefont {G.}~\bibnamefont {{de la Flor}}}, \ and\ \bibinfo
  {author} {\bibfnamefont {M.~I.}\ \bibnamefont {Aroyo}},\ }\bibfield  {title}
  {\enquote {\bibinfo {title} {Symmetry-based computational tools for magnetic
  crystallography},}\ }\href {\doibase 10.1146/annurev-matsci-070214-021008}
  {\bibfield  {journal} {\bibinfo  {journal} {Annu. Rev. Mater. Res.}\ }\textbf
  {\bibinfo {volume} {45}},\ \bibinfo {pages} {217} (\bibinfo {year}
  {2015})}\BibitemShut {NoStop}%
\bibitem [{\citenamefont {Bertin}\ \emph {et~al.}(2015)\citenamefont {Bertin},
  \citenamefont {Dalmas~de R\'eotier}, \citenamefont {F\aa{}k}, \citenamefont
  {Marin}, \citenamefont {Yaouanc}, \citenamefont {Forget}, \citenamefont
  {Sheptyakov}, \citenamefont {Frick}, \citenamefont {Ritter}, \citenamefont
  {Amato}, \citenamefont {Baines},\ and\ \citenamefont {King}}]{Bertin15}%
  \BibitemOpen
  \bibfield  {author} {\bibinfo {author} {\bibfnamefont {A.}~\bibnamefont
  {Bertin}}, \bibinfo {author} {\bibfnamefont {P.}~\bibnamefont {Dalmas~de
  R\'eotier}}, \bibinfo {author} {\bibfnamefont {B.}~\bibnamefont {F\aa{}k}},
  \bibinfo {author} {\bibfnamefont {C.}~\bibnamefont {Marin}}, \bibinfo
  {author} {\bibfnamefont {A.}~\bibnamefont {Yaouanc}}, \bibinfo {author}
  {\bibfnamefont {A.}~\bibnamefont {Forget}}, \bibinfo {author} {\bibfnamefont
  {D.}~\bibnamefont {Sheptyakov}}, \bibinfo {author} {\bibfnamefont
  {B.}~\bibnamefont {Frick}}, \bibinfo {author} {\bibfnamefont
  {C.}~\bibnamefont {Ritter}}, \bibinfo {author} {\bibfnamefont
  {A.}~\bibnamefont {Amato}}, \bibinfo {author} {\bibfnamefont
  {C.}~\bibnamefont {Baines}}, \ and\ \bibinfo {author} {\bibfnamefont
  {P.~J.~C.}\ \bibnamefont {King}},\ }\bibfield  {title} {\enquote {\bibinfo
  {title} {$\mathrm{Nd}_{2}\mathrm{Sn}_{2}\mathrm{O}_{7}$: An all-in--all-out
  pyrochlore magnet with no divergence-free field and anomalously slow
  paramagnetic spin dynamics},}\ }\href {\doibase 10.1103/PhysRevB.92.144423}
  {\bibfield  {journal} {\bibinfo  {journal} {Phys. Rev. B}\ }\textbf {\bibinfo
  {volume} {92}},\ \bibinfo {pages} {144423} (\bibinfo {year}
  {2015})}\BibitemShut {NoStop}%
\bibitem [{\citenamefont {Mirebeau}\ \emph {et~al.}(2005)\citenamefont
  {Mirebeau}, \citenamefont {Apetrei}, \citenamefont
  {{Rodr{\'\i}guez-Carvajal}}, \citenamefont {Bonville}, \citenamefont
  {Forget}, \citenamefont {Colson}, \citenamefont {Glazkov}, \citenamefont
  {Sanchez}, \citenamefont {Isnard},\ and\ \citenamefont {Suard}}]{Mirebeau05}%
  \BibitemOpen
  \bibfield  {author} {\bibinfo {author} {\bibfnamefont {I.}~\bibnamefont
  {Mirebeau}}, \bibinfo {author} {\bibfnamefont {A.}~\bibnamefont {Apetrei}},
  \bibinfo {author} {\bibfnamefont {J.}~\bibnamefont
  {{Rodr{\'\i}guez-Carvajal}}}, \bibinfo {author} {\bibfnamefont
  {P.}~\bibnamefont {Bonville}}, \bibinfo {author} {\bibfnamefont
  {A.}~\bibnamefont {Forget}}, \bibinfo {author} {\bibfnamefont
  {D.}~\bibnamefont {Colson}}, \bibinfo {author} {\bibfnamefont
  {V.}~\bibnamefont {Glazkov}}, \bibinfo {author} {\bibfnamefont {J.~P.}\
  \bibnamefont {Sanchez}}, \bibinfo {author} {\bibfnamefont {O.}~\bibnamefont
  {Isnard}}, \ and\ \bibinfo {author} {\bibfnamefont {E.}~\bibnamefont
  {Suard}},\ }\bibfield  {title} {\enquote {\bibinfo {title} {Ordered spin ice
  state and magnetic fluctuations in
  \uppercase{T}b$_2$\uppercase{S}n$_2$\uppercase{O}$_7$},}\ }\href {\doibase
  10.1103/PhysRevLett.94.246402} {\bibfield  {journal} {\bibinfo  {journal}
  {Phys. Rev. Lett.}\ }\textbf {\bibinfo {volume} {94}},\ \bibinfo {pages}
  {246402} (\bibinfo {year} {2005})}\BibitemShut {NoStop}%
\bibitem [{\citenamefont {{Dalmas de R\'eotier}}\ \emph
  {et~al.}(2006)\citenamefont {{Dalmas de R\'eotier}}, \citenamefont {Yaouanc},
  \citenamefont {Keller}, \citenamefont {Cervellino}, \citenamefont {Roessli},
  \citenamefont {Baines}, \citenamefont {Forget}, \citenamefont {Vaju},
  \citenamefont {Gubbens}, \citenamefont {Amato},\ and\ \citenamefont
  {King}}]{Dalmas06}%
  \BibitemOpen
  \bibfield  {author} {\bibinfo {author} {\bibfnamefont {P.}~\bibnamefont
  {{Dalmas de R\'eotier}}}, \bibinfo {author} {\bibfnamefont {A.}~\bibnamefont
  {Yaouanc}}, \bibinfo {author} {\bibfnamefont {L.}~\bibnamefont {Keller}},
  \bibinfo {author} {\bibfnamefont {A.}~\bibnamefont {Cervellino}}, \bibinfo
  {author} {\bibfnamefont {B.}~\bibnamefont {Roessli}}, \bibinfo {author}
  {\bibfnamefont {C.}~\bibnamefont {Baines}}, \bibinfo {author} {\bibfnamefont
  {A.}~\bibnamefont {Forget}}, \bibinfo {author} {\bibfnamefont
  {C.}~\bibnamefont {Vaju}}, \bibinfo {author} {\bibfnamefont {P.~C.~M.}\
  \bibnamefont {Gubbens}}, \bibinfo {author} {\bibfnamefont {A.}~\bibnamefont
  {Amato}}, \ and\ \bibinfo {author} {\bibfnamefont {P.~J.~C.}\ \bibnamefont
  {King}},\ }\bibfield  {title} {\enquote {\bibinfo {title} {Spin dynamics and
  magnetic order in magnetically frustrated
  \uppercase{T}b$_2$\uppercase{S}n$_2$\uppercase{O}$_7$},}\ }\href {\doibase
  10.1103/PhysRevLett.96.127202} {\bibfield  {journal} {\bibinfo  {journal}
  {Phys. Rev. Lett.}\ }\textbf {\bibinfo {volume} {96}},\ \bibinfo {pages}
  {127202} (\bibinfo {year} {2006})}\BibitemShut {NoStop}%
\bibitem [{\citenamefont {Lago}\ \emph {et~al.}(2014)\citenamefont {Lago},
  \citenamefont {\ifmmode \check{Z}\else
  \v{Z}\fi{}ivkovi\ifmmode~\acute{c}\else \'{c}\fi{}}, \citenamefont {Piatek},
  \citenamefont {\'Alvarez}, \citenamefont {H\"uvonen}, \citenamefont {Pratt},
  \citenamefont {{D\'\i az}},\ and\ \citenamefont {Rojo}}]{Lago14}%
  \BibitemOpen
  \bibfield  {author} {\bibinfo {author} {\bibfnamefont {J.}~\bibnamefont
  {Lago}}, \bibinfo {author} {\bibfnamefont {I.}~\bibnamefont {\ifmmode
  \check{Z}\else \v{Z}\fi{}ivkovi\ifmmode~\acute{c}\else \'{c}\fi{}}}, \bibinfo
  {author} {\bibfnamefont {J.~O.}\ \bibnamefont {Piatek}}, \bibinfo {author}
  {\bibfnamefont {P.}~\bibnamefont {\'Alvarez}}, \bibinfo {author}
  {\bibfnamefont {D.}~\bibnamefont {H\"uvonen}}, \bibinfo {author}
  {\bibfnamefont {F.~L.}\ \bibnamefont {Pratt}}, \bibinfo {author}
  {\bibfnamefont {M.}~\bibnamefont {{D\'\i az}}}, \ and\ \bibinfo {author}
  {\bibfnamefont {T.}~\bibnamefont {Rojo}},\ }\bibfield  {title} {\enquote
  {\bibinfo {title} {Glassy dynamics in the low-temperature inhomogeneous
  ferromagnetic phase of the quantum spin ice
  \uppercase{Y}b$_2$\uppercase{S}n$_2$\uppercase{O}$_7$},}\ }\href {\doibase
  10.1103/PhysRevB.89.024421} {\bibfield  {journal} {\bibinfo  {journal} {Phys.
  Rev. B}\ }\textbf {\bibinfo {volume} {89}},\ \bibinfo {pages} {024421}
  (\bibinfo {year} {2014})}\BibitemShut {NoStop}%
\bibitem [{Note2()}]{Note2}%
  \BibitemOpen
  \bibinfo {note} {The pyrochlore Gd$_2$Ti$_2$O$_7$ is not listed because its
  propagation wavevector is ${\protect \bf k}= (1/2,1/2,1/2)$. In fact, its
  magnetic structure is still under debate \cite {Paddison15}.}\BibitemShut
  {Stop}%
\bibitem [{\citenamefont {Yan}\ \emph {et~al.}(2017)\citenamefont {Yan},
  \citenamefont {Benton}, \citenamefont {Jaubert},\ and\ \citenamefont
  {Shannon}}]{Yan17}%
  \BibitemOpen
  \bibfield  {author} {\bibinfo {author} {\bibfnamefont {Han}\ \bibnamefont
  {Yan}}, \bibinfo {author} {\bibfnamefont {Owen}\ \bibnamefont {Benton}},
  \bibinfo {author} {\bibfnamefont {Ludovic}\ \bibnamefont {Jaubert}}, \ and\
  \bibinfo {author} {\bibfnamefont {Nic}\ \bibnamefont {Shannon}},\ }\bibfield
  {title} {\enquote {\bibinfo {title} {Theory of multiple-phase competition in
  pyrochlore magnets with anisotropic exchange with application to
  $\mathrm{Yb}_{2}\mathrm{Ti}_{2}\mathrm{O}_{7},
  \mathrm{Er}_{2}\mathrm{Ti}_{2}\mathrm{O}_{7}$, and
  $\mathrm{Er}_{2}\mathrm{Sn}_{2}\mathrm{O}_{7}$},}\ }\href {\doibase
  10.1103/PhysRevB.95.094422} {\bibfield  {journal} {\bibinfo  {journal} {Phys.
  Rev. B}\ }\textbf {\bibinfo {volume} {95}},\ \bibinfo {pages} {094422}
  (\bibinfo {year} {2017})}\BibitemShut {NoStop}%
\bibitem [{\citenamefont {Curnoe}(2008)}]{Curnoe08}%
  \BibitemOpen
  \bibfield  {author} {\bibinfo {author} {\bibfnamefont {S.~H.}\ \bibnamefont
  {Curnoe}},\ }\bibfield  {title} {\enquote {\bibinfo {title} {Structural
  distortion and the spin liquid state in
  $\text{Tb}_{2}\text{Ti}_{2}\text{O}_{7}$},}\ }\href {\doibase
  10.1103/PhysRevB.78.094418} {\bibfield  {journal} {\bibinfo  {journal} {Phys.
  Rev. B}\ }\textbf {\bibinfo {volume} {78}},\ \bibinfo {pages} {094418}
  (\bibinfo {year} {2008})}\BibitemShut {NoStop}%
\bibitem [{\citenamefont {Hartmann}\ \emph {et~al.}(2013)\citenamefont
  {Hartmann}, \citenamefont {Kalvius}, \citenamefont {W\"appling},
  \citenamefont {G\"unther}, \citenamefont {Tsurkan}, \citenamefont {Krimmel},\
  and\ \citenamefont {Loid}}]{Hartmann13}%
  \BibitemOpen
  \bibfield  {author} {\bibinfo {author} {\bibfnamefont {Ola}\ \bibnamefont
  {Hartmann}}, \bibinfo {author} {\bibfnamefont {Georg~Michael}\ \bibnamefont
  {Kalvius}}, \bibinfo {author} {\bibfnamefont {Roger}\ \bibnamefont
  {W\"appling}}, \bibinfo {author} {\bibfnamefont {Axel}\ \bibnamefont
  {G\"unther}}, \bibinfo {author} {\bibfnamefont {Vladimir}\ \bibnamefont
  {Tsurkan}}, \bibinfo {author} {\bibfnamefont {Alex}\ \bibnamefont {Krimmel}},
  \ and\ \bibinfo {author} {\bibfnamefont {Alois}\ \bibnamefont {Loid}},\
  }\bibfield  {title} {\enquote {\bibinfo {title} {Magnetic properties of the
  multiferroic chromium thio-spinels
  \uppercase{C}d\uppercase{C}r$_2$\uppercase{S}$_4$ and
  \uppercase{H}g\uppercase{C}r$_2$\uppercase{S}$_4$},}\ }\href {\doibase
  doi:10.1140/epjb/e2013-31094-4} {\bibfield  {journal} {\bibinfo  {journal}
  {Eur. Phys. J. B}\ }\textbf {\bibinfo {volume} {86}},\ \bibinfo {pages} {148}
  (\bibinfo {year} {2013})}\BibitemShut {NoStop}%
\bibitem [{\citenamefont {Baltzer}\ \emph {et~al.}(1965)\citenamefont
  {Baltzer}, \citenamefont {Lehmann},\ and\ \citenamefont
  {Robbins}}]{Baltzer65}%
  \BibitemOpen
  \bibfield  {author} {\bibinfo {author} {\bibfnamefont {P.~K.}\ \bibnamefont
  {Baltzer}}, \bibinfo {author} {\bibfnamefont {H.~W.}\ \bibnamefont
  {Lehmann}}, \ and\ \bibinfo {author} {\bibfnamefont {M.}~\bibnamefont
  {Robbins}},\ }\bibfield  {title} {\enquote {\bibinfo {title} {Insulating
  ferromagnetic spinels},}\ }\href {\doibase 10.1103/PhysRevLett.15.493}
  {\bibfield  {journal} {\bibinfo  {journal} {Phys. Rev. Lett.}\ }\textbf
  {\bibinfo {volume} {15}},\ \bibinfo {pages} {493} (\bibinfo {year}
  {1965})}\BibitemShut {NoStop}%
\bibitem [{\citenamefont {Brooks-Bartlett}\ \emph {et~al.}(2014)\citenamefont
  {Brooks-Bartlett}, \citenamefont {Banks}, \citenamefont {Jaubert},
  \citenamefont {Harman-Clarke},\ and\ \citenamefont {Holdsworth}}]{Brooks14}%
  \BibitemOpen
  \bibfield  {author} {\bibinfo {author} {\bibfnamefont {M.~E.}\ \bibnamefont
  {Brooks-Bartlett}}, \bibinfo {author} {\bibfnamefont {S.~T.}\ \bibnamefont
  {Banks}}, \bibinfo {author} {\bibfnamefont {L.~D.~C.}\ \bibnamefont
  {Jaubert}}, \bibinfo {author} {\bibfnamefont {A.}~\bibnamefont
  {Harman-Clarke}}, \ and\ \bibinfo {author} {\bibfnamefont {P.~C.~W.}\
  \bibnamefont {Holdsworth}},\ }\bibfield  {title} {\enquote {\bibinfo {title}
  {Magnetic-moment fragmentation and monopole crystallization},}\ }\href
  {\doibase 10.1103/PhysRevX.4.011007} {\bibfield  {journal} {\bibinfo
  {journal} {Phys. Rev. X}\ }\textbf {\bibinfo {volume} {4}},\ \bibinfo {pages}
  {011007} (\bibinfo {year} {2014})}\BibitemShut {NoStop}%
\bibitem [{\citenamefont {Orend\'a\ifmmode~\check{c}\else \v{c}\fi{}}\ \emph
  {et~al.}(2016)\citenamefont {Orend\'a\ifmmode~\check{c}\else \v{c}\fi{}},
  \citenamefont {Tibensk\'a}, \citenamefont {Stre\ifmmode~\check{c}\else
  \v{c}\fi{}ka}, \citenamefont {\ifmmode~\check{C}\else \v{C}\fi{}is\'arov\'a},
  \citenamefont {Tk\'a\ifmmode~\check{c}\else \v{c}\fi{}}, \citenamefont
  {Orend\'a\ifmmode~\check{c}\else \v{c}\fi{}ov\'a}, \citenamefont {\ifmmode
  \check{C}\else \v{C}\fi{}i\ifmmode~\check{z}\else \v{z}\fi{}m\'ar},
  \citenamefont {Prokle\ifmmode~\check{s}\else \v{s}\fi{}ka},\ and\
  \citenamefont {Sechovsk\'y}}]{Orendac16}%
  \BibitemOpen
  \bibfield  {author} {\bibinfo {author} {\bibfnamefont {M.}~\bibnamefont
  {Orend\'a\ifmmode~\check{c}\else \v{c}\fi{}}}, \bibinfo {author}
  {\bibfnamefont {K.}~\bibnamefont {Tibensk\'a}}, \bibinfo {author}
  {\bibfnamefont {J.}~\bibnamefont {Stre\ifmmode~\check{c}\else \v{c}\fi{}ka}},
  \bibinfo {author} {\bibfnamefont {J.}~\bibnamefont {\ifmmode~\check{C}\else
  \v{C}\fi{}is\'arov\'a}}, \bibinfo {author} {\bibfnamefont {V.}~\bibnamefont
  {Tk\'a\ifmmode~\check{c}\else \v{c}\fi{}}}, \bibinfo {author} {\bibfnamefont
  {A.}~\bibnamefont {Orend\'a\ifmmode~\check{c}\else \v{c}\fi{}ov\'a}},
  \bibinfo {author} {\bibfnamefont {E.}~\bibnamefont {\ifmmode \check{C}\else
  \v{C}\fi{}i\ifmmode~\check{z}\else \v{z}\fi{}m\'ar}}, \bibinfo {author}
  {\bibfnamefont {J.}~\bibnamefont {Prokle\ifmmode~\check{s}\else
  \v{s}\fi{}ka}}, \ and\ \bibinfo {author} {\bibfnamefont {V.}~\bibnamefont
  {Sechovsk\'y}},\ }\bibfield  {title} {\enquote {\bibinfo {title}
  {Cross-tunneling and phonon bottleneck effects in the relaxation phenomena of
  $\uppercase{XY}$ pyrochlore antiferromagnet
  \uppercase{E}r$_2$\uppercase{T}i$_2$\uppercase{O}$_7$},}\ }\href {\doibase
  10.1103/PhysRevB.93.024410} {\bibfield  {journal} {\bibinfo  {journal} {Phys.
  Rev. B}\ }\textbf {\bibinfo {volume} {93}},\ \bibinfo {pages} {024410}
  (\bibinfo {year} {2016})}\BibitemShut {NoStop}%
\bibitem [{\citenamefont {Bloembergen}\ \emph {et~al.}(1959)\citenamefont
  {Bloembergen}, \citenamefont {Shapiro}, \citenamefont {Pershan},\ and\
  \citenamefont {Artman}}]{Bloembergen59}%
  \BibitemOpen
  \bibfield  {author} {\bibinfo {author} {\bibfnamefont {N.}~\bibnamefont
  {Bloembergen}}, \bibinfo {author} {\bibfnamefont {S.}~\bibnamefont
  {Shapiro}}, \bibinfo {author} {\bibfnamefont {P.~S.}\ \bibnamefont
  {Pershan}}, \ and\ \bibinfo {author} {\bibfnamefont {J.~O.}\ \bibnamefont
  {Artman}},\ }\bibfield  {title} {\enquote {\bibinfo {title} {Cross-relaxation
  in spin systems},}\ }\href {\doibase 10.1103/PhysRev.114.445} {\bibfield
  {journal} {\bibinfo  {journal} {Phys. Rev.}\ }\textbf {\bibinfo {volume}
  {114}},\ \bibinfo {pages} {445} (\bibinfo {year} {1959})}\BibitemShut
  {NoStop}%
\bibitem [{\citenamefont {Ruminy}\ \emph {et~al.}(2016)\citenamefont {Ruminy},
  \citenamefont {Pomjakushina}, \citenamefont {Iida}, \citenamefont {Kamazawa},
  \citenamefont {Adroja}, \citenamefont {Stuhr},\ and\ \citenamefont
  {Fennell}}]{Ruminy16}%
  \BibitemOpen
  \bibfield  {author} {\bibinfo {author} {\bibfnamefont {M.}~\bibnamefont
  {Ruminy}}, \bibinfo {author} {\bibfnamefont {E.}~\bibnamefont
  {Pomjakushina}}, \bibinfo {author} {\bibfnamefont {K.}~\bibnamefont {Iida}},
  \bibinfo {author} {\bibfnamefont {K.}~\bibnamefont {Kamazawa}}, \bibinfo
  {author} {\bibfnamefont {D.~T.}\ \bibnamefont {Adroja}}, \bibinfo {author}
  {\bibfnamefont {U.}~\bibnamefont {Stuhr}}, \ and\ \bibinfo {author}
  {\bibfnamefont {T.}~\bibnamefont {Fennell}},\ }\bibfield  {title} {\enquote
  {\bibinfo {title} {Crystal-field parameters of the rare-earth pyrochlores
  $\uppercase{R}$$_2$\uppercase{T}i$_2$\uppercase{O}$_7$ ($\uppercase{R} = $
  \uppercase{T}b, \uppercase{D}y, and \uppercase{H})},}\ }\href {\doibase
  10.1103/PhysRevB.94.024430} {\bibfield  {journal} {\bibinfo  {journal} {Phys.
  Rev. B}\ }\textbf {\bibinfo {volume} {94}},\ \bibinfo {pages} {024430}
  (\bibinfo {year} {2016})}\BibitemShut {NoStop}%
\bibitem [{\citenamefont {Yoshizawa}\ \emph {et~al.}(2015)\citenamefont
  {Yoshizawa}, \citenamefont {Kida}, \citenamefont {Nakatsuji}, \citenamefont
  {Iritani}, \citenamefont {Halim}, \citenamefont {Takeuchi},\ and\
  \citenamefont {Hagiwara}}]{Yoshizawa15}%
  \BibitemOpen
  \bibfield  {author} {\bibinfo {author} {\bibfnamefont {D.}~\bibnamefont
  {Yoshizawa}}, \bibinfo {author} {\bibfnamefont {T.}~\bibnamefont {Kida}},
  \bibinfo {author} {\bibfnamefont {S.}~\bibnamefont {Nakatsuji}}, \bibinfo
  {author} {\bibfnamefont {K.}~\bibnamefont {Iritani}}, \bibinfo {author}
  {\bibfnamefont {M.}~\bibnamefont {Halim}}, \bibinfo {author} {\bibfnamefont
  {T.}~\bibnamefont {Takeuchi}}, \ and\ \bibinfo {author} {\bibfnamefont
  {M.}~\bibnamefont {Hagiwara}},\ }\bibfield  {title} {\enquote {\bibinfo
  {title} {High-field multi-frequency \uppercase{ESR} in the rare-earth spinel
  compound \uppercase{C}d\uppercase{Y}b$_2$\uppercase{S}$_4$},}\ }\href
  {\doibase 10.1007/s00723-015-0651-x} {\bibfield  {journal} {\bibinfo
  {journal} {Appl. Magn. Reson.}\ }\textbf {\bibinfo {volume} {46}},\ \bibinfo
  {pages} {993} (\bibinfo {year} {2015})}\BibitemShut {NoStop}%
\bibitem [{\citenamefont {Gaudet}\ \emph {et~al.}(2016)\citenamefont {Gaudet},
  \citenamefont {Ross}, \citenamefont {Kermarrec}, \citenamefont {Butch},
  \citenamefont {Ehlers}, \citenamefont {Dabkowska},\ and\ \citenamefont
  {Gaulin}}]{Gaudet16}%
  \BibitemOpen
  \bibfield  {author} {\bibinfo {author} {\bibfnamefont {J.}~\bibnamefont
  {Gaudet}}, \bibinfo {author} {\bibfnamefont {K.~A.}\ \bibnamefont {Ross}},
  \bibinfo {author} {\bibfnamefont {E.}~\bibnamefont {Kermarrec}}, \bibinfo
  {author} {\bibfnamefont {N.~P.}\ \bibnamefont {Butch}}, \bibinfo {author}
  {\bibfnamefont {G.}~\bibnamefont {Ehlers}}, \bibinfo {author} {\bibfnamefont
  {H.~A.}\ \bibnamefont {Dabkowska}}, \ and\ \bibinfo {author} {\bibfnamefont
  {B.~D.}\ \bibnamefont {Gaulin}},\ }\bibfield  {title} {\enquote {\bibinfo
  {title} {Gapless quantum excitations from an icelike splayed ferromagnetic
  ground state in stoichiometric
  \uppercase{Y}b$_2$\uppercase{T}i$_2$\uppercase{O}$_7$},}\ }\href {\doibase
  10.1103/PhysRevB.93.064406} {\bibfield  {journal} {\bibinfo  {journal} {Phys.
  Rev. B}\ }\textbf {\bibinfo {volume} {93}},\ \bibinfo {pages} {064406}
  (\bibinfo {year} {2016})}\BibitemShut {NoStop}%
\bibitem [{\citenamefont {Paddison}\ \emph {et~al.}(2015)\citenamefont
  {Paddison}, \citenamefont {Cairns}, \citenamefont {Khalyavin}, \citenamefont
  {Manuel}, \citenamefont {Daoud-Aladine}, \citenamefont {Ehlers},
  \citenamefont {Petrenko}, \citenamefont {Gardner}, \citenamefont {Zhou},
  \citenamefont {Goodwin},\ and\ \citenamefont {Stewart}}]{Paddison15}%
  \BibitemOpen
  \bibfield  {author} {\bibinfo {author} {\bibfnamefont {Joseph A.~M.}\
  \bibnamefont {Paddison}}, \bibinfo {author} {\bibfnamefont {Andrew~B.}\
  \bibnamefont {Cairns}}, \bibinfo {author} {\bibfnamefont {Dmitry~D.}\
  \bibnamefont {Khalyavin}}, \bibinfo {author} {\bibfnamefont {Pascal}\
  \bibnamefont {Manuel}}, \bibinfo {author} {\bibfnamefont {Aziz}\ \bibnamefont
  {Daoud-Aladine}}, \bibinfo {author} {\bibfnamefont {Georg}\ \bibnamefont
  {Ehlers}}, \bibinfo {author} {\bibfnamefont {Oleg~A.}\ \bibnamefont
  {Petrenko}}, \bibinfo {author} {\bibfnamefont {Jason~S.}\ \bibnamefont
  {Gardner}}, \bibinfo {author} {\bibfnamefont {H.~D.}\ \bibnamefont {Zhou}},
  \bibinfo {author} {\bibfnamefont {Andrew~L.}\ \bibnamefont {Goodwin}}, \ and\
  \bibinfo {author} {\bibfnamefont {J.~Ross}\ \bibnamefont {Stewart}},\
  }\href@noop {} {\enquote {\bibinfo {title} {Nature of partial magnetic order
  in the frustrated antiferromagnet
  \uppercase{G}d$_2$\uppercase{T}i$_2$\uppercase{O}$_7$},}\ } (\bibinfo {year}
  {2015}),\ \bibinfo {note} {arXiv:1506.05045}\BibitemShut {NoStop}%
\end{thebibliography}%

\end{document}